\documentclass[%
 reprint,
 amsmath,amssymb,
 aps,
]{emulateapj}

\begin{document}

% Title of the paper, and the short title which is used in the headers.
% Keep the title short and informative.
\title[The Vanishing \& Appearing Sources during a Century of Observations project]{The Vanishing \& Appearing Sources during a Century of Observations project:\\
I. USNO objects missing in modern sky surveys and follow-up observations of a ``missing star''}

% The list of authors, and the short list which is used in the headers.
% If you need two or more lines of authors, add an extra line using \newauthor
\author{Beatriz Villarroel}
\affiliation{Nordita, KTH Royal Institute of Technology and Stockholm University, Roslagstullsbacken 23, SE-106 91 Stockholm, Sweden\\
Instituto de Astrofisica de Canarias, Avda Via Lactea S/N, La Laguna, E-38205, Tenerife, Spain}
\author{Johan Soodla}
\affiliation{Department of Information Technology, Uppsala University, Uppsala, Sweden}
\author{S\'{e}bastien Comer\'{o}n}
\affiliation{University of Oulu, Astronomy Research Unit, 90014 Oulu, Finland}
\author{Lars Mattsson}
\affiliation{Nordita, KTH Royal Institute of Technology and Stockholm University, Roslagstullsbacken 23, SE-106 91 Stockholm, Sweden}
\author{Kristiaan Pelckmans}
\affiliation{Department of Information Technology, Uppsala University, Uppsala, Sweden}
\author{Mart\'{i}n L\'{o}pez Corredoira}
\affiliation{Instituto de Astrofisica de Canarias, Avda Via Lactea S/N, La Laguna, E-38205, Tenerife, Spain\\
Departamento de Astrofisica, Universidad de La Laguna, E-38206 La Laguna, Tenerife, Spain}
\author{Kevin Krisciunas}
\affiliation{Department of Physics and Astronomy, Texas A\&M University, 4242 TAMU, College
Station, Texas 77843, USA}
\author{Eduardo Guerras}
\affiliation{Homer L. Dodge Department of Physics and Astronomy, The University of Oklahoma, Norman, OK, 73019, USA}
\author{Oleg Kochukhov}
\affiliation{ Department of Physics and Astronomy, Uppsala University, Uppsala, Sweden}
\author{Josefine Bergstedt}
\affiliation{ Department of Physics and Astronomy, Uppsala University, Uppsala, Sweden}
\author{Bart Buelens}
\affiliation{Statistics Netherlands, Methodology Department, Heerlen, The Netherlands\\
VITO, Diepenbeek Genk, Flanders, Belgium}
\author{Rudolf E. B\"ar}
\affiliation{Institute for Particle Physics and Astrophysics, ETH Zurich, Wolfgang-Pauli-Strasse 27, CH-8093 Zurich, Switzerland}
\author{Rub\'{e}n Cubo}
\affiliation{Department of Information Technology, Uppsala University, Uppsala, Sweden}
\author{J. Emilio Enriquez}
\affiliation{Department of Astronomy, University of California Berkeley, 501 Campbell Hall \#311, Berkeley, CA, 94720}
\author{Alok C. Gupta}
\affiliation{Aryabhatta Research Institute of Observational Sciences (ARIES), Manora Peak, Nainital, 263 002, India}
\author{I\~{n}igo Imaz}
\affiliation{ Department of Physics and Astronomy, Uppsala University, Uppsala, Sweden}
\author{Torgny Karlsson}
\affiliation{Department of Immunology, Uppsala University, Uppsala, Sweden}
\author{M. Almudena Prieto}
\affiliation{Instituto de Astrofisica de Canarias, Avda Via Lactea S/N, La Laguna, E-38205, Tenerife, Spain\\
Departamento de Astrofisica, Universidad de La Laguna, E-38206 La Laguna, Tenerife, Spain}
\author{Aleksey A. Shlyapnikov}
\affiliation{Stellar Physics Department, Crimean Astrophysical Observatory, 298409, Nauchnyj, Crimea}
\author{Rafael S. de Souza}
\affiliation{Department of Physics \& Astronomy, University of North Carolina at Chapel Hill, NC 27599-3255, USA}
\author{Irina B. Vavilova}
\affiliation{Main Astronomical Observatory of the NAS of Ukraine, 27, Akademik Zabolotny St., Kyiv, 03143, Ukraine}
\author{Martin J. Ward}
\affiliation{Centre for Extragalactic Astronomy, Department of Physics, Durham University, South Road, Durham, DH1 3LE, UK}

\label{firstpage}
\newpage
%\thanks{E-mail: beatriz.villarroel@su.se}
% These dates will be filled out by the publisher
%\date{Accepted XXX. Received YYY; in original form ZZZ}

% Enter the current year, for the copyright statements etc.

%\date{\today}

\begin{abstract}
In this paper we report the current status of a new research program. The primary goal of the ``Vanishing \& Appearing Sources during a Century of Observations'' (VASCO) project is to search for vanishing and appearing sources using existing survey data to find examples of exceptional astrophysical transients. The implications of finding such objects extend from traditional astrophysics fields to the more exotic searches for evidence of technologically advanced civilizations. In this first paper we present new, deeper observations of the tentative candidate discovered by \cite{Villarroel2016}. We then perform the first searches for vanishing objects throughout the sky by comparing 600 million objects from the US Naval Observatory Catalogue (USNO) B1.0 down to a limiting magnitude of $\sim 20 - 21$ with the recent Pan-STARRS Data Release-1 (DR1) with a limiting magnitude of $\sim$ 23.4. We find about 150,000 preliminary candidates that do not have any Pan-STARRS counterpart within a 30 arcsec radius. We show that these objects are redder and have larger proper motions than typical USNO objects. We visually examine the images for a subset of about 24,000 candidates, superseding the 2016 study with a sample ten times larger. We find about $\sim$ 100 point sources visible in only one epoch in the red band of the USNO which may be of interest in searches for strong M dwarf flares, high-redshift supernovae or other catagories of unidentified red transients. %%We estimate that the chance of finding ``vanishing'' stars in the Milky Way is less than 1 in 90 million during 70 years. The VASCO project will likely generate new lists of interesting events.
\end{abstract}


%%%%%%%%%%%%%%%%%%%%%%%%%%%%%%%%%%%%%%%%%%%%%%%%%%

%%%%%%%%%%%%%%%%% BODY OF PAPER %%%%%%%%%%%%%%%%%%

\section{Introduction}\label{sec:intro}
%%{\it When you look into an abyss, the abyss also looks into you.}\\
%%\\

Many of the hottest topics in current astronomical research concern the physics of extreme transient 
phenomena, such as gravitational wave events, gamma-ray bursts, Fast Radio Bursts (FRBs) or Active Galactic Nuclei (AGN) outbursts. Although we are gaining a better understanding of the physical processes governing them, our understanding of the transient phenomena in general is inevitably limited by the {\em a priori} assumptions that go into the data collection when we design our observations. With the advent of the Virtual Observatory in the early 2000s,
astronomers suggested that very large surveys together with state-of-the-art 
developments in information technology could efficiently be used 
to probe rare or unusual astrophysical phenomena by expanding the parameter space beyond our current knowledge \citep{Djorgovski2000,Djorgovski2001}. An example of such a rare class of object that would not have been discovered unless specifically looked for, is that of Hippke's star. This emerged from a search for artificially modified pulsations in Cepheid variables and led to the discovery of rare objects with two regimes with both long and short duration double pulsation periods \citep{Hippke2015}. 

Another example of objects that may be missed in transient surveys, unless specifically looked for, are the rare failed supernovae \citep{Kochanek}, which occur when a star collapses almost directly to form a black hole. Recently, the possible detection of a failed supernova in a nearby galaxy has been reported \citep{Adams2017a,Adams2017b}. There more exotic the phenomenon, the more likely are to miss it in the observational data due to our preconceptions and the duration and frequency of the sampling.

%In this paper, we launch the ``Vanishing \& Appearing Sources during a Century of Observations'' (VASCO) project, a cross-disciplinary effort aimed at finding some of the most unusual variable phenomena and other astrophysical anomalies with help of existing large surveys, machine learning and citizen science. The project is 
In this paper we describe the ``Vanishing \& Appearing Sources during a Century of Observations'' (VASCO) project\footnote{https://vasconsite.wordpress.com}, a multitask  effort aimed at finding some of the most unusual variable phenomena and other astrophysical anomalies based on existing sky surveys. We also aim to develop a citizen-science branch of VASCO and indeed the basic philosophy behind the project was first described for a wider audience by \citet{Mattsson2017}. 

VASCO is primarily centered around searches for vanishing objects observed in the sky and beyond the Earth's local environment. Unless a star directly collapses into a black hole, there is no known physical process by which it could physically vanish. If such examples exist this makes it interesting for searches for new exotic phenomena or even signs of technologically advanced civilisations \citep{Villarroel2016}. Vanishing stellar events currently are missed and hence go undetected in most ongoing all-sky surveys.  \cite{Villarroel2016} found only one such tentative candidate after a cross-match between 10 million USNO sources and the SDSS. Even if we discover a star that appears to vanish, it is an observational challenge to determine whether the object really vanished or just faded below the detection limit.

The VASCO project aims to find both vanishing and appearing sources as well as 
objects that show extreme variability on extended time scales (many decades), by comparing the nearly a century-old ($\sim$ 70 years) sky scans with modern-day astronomical surveys. Compared to the recent transient facilities such as the Zwicky Transient Facility (ZTF) that commenced operations in 2018, we are probing a significantly longer time window -- about 70 years -- by  
investigating events that occurred between the epoch of the US Naval Observatory Catalogue catalogue \citep{Monet2003} and the recent Pan-STARRS survey that has multiple detections for each astronomical source \citep{Kaiser2002}. 
Prior efforts to probe these large timescales have been led by the ``Digital Access to a Sky Century @ Harvard'' (DASCH) project \citep{Grindlay2012} that has digitized more than 450,000 plates with a full-sky coverage. The plates used were taken during the years 1890 to 1990 and had a limiting magnitude of $B \sim$ 14 (or $V \sim$ 15). Among the modern CCD surveys the Catalina Real-Time Transient Survey (CRTS) has the largest time span ($\sim$ 14 years), with a total sky coverage of about 30,000 deg$^{2}$ and about ~500 million light curves \citep{Drake2009,Mahabal,Djorgovski2012}. The data, which is public, extends to $V \sim 19 - 21$ mag per exposure and is based on CCD photometry taken at a large number of epochs. The CRTS has so far discovered about $\sim$ 17,000 optical transients, among them many superluminous and pecular supernovae, about 1500 cataclysmic variables, and about 4000 variable AGN.

Using a relatively large time window of $\sim$ 70 years, in combination with a large sample size, increases the probability of finding extremely rare events. Clearly this is still a minute time duration from a cosmological perspective, but it nevertheless sets an upper limit on the incidence of vanishing or appearing-star events. In addition there are a number of recently discovered astronomical transients that occur over significantly longer time scales than are for common variable stars that 
vary on periods from weeks to a few years. For example, hypervariable AGN \citep{Lawrence,Kankare2017} were discovered by comparing two astronomical surveys separated by a ten year time gap. More than 95 percent of extragalactic objects exhibiting this long-term variability show the presence of an AGN \citep{Drake2019}. Hypervariable AGN exhibit still poorly understood long-term variability that could have various causes, e.g. microlensing events, superluminous supernovae in the accretion disk \citep{Graham2017} or changes in the Eddington ratio of the AGN \citep{Graham2019}. These hypervariable AGN have been extensively studied with the CRTS.

The DASCH project has reported other interesting findings while probing these timescales. For example, it revealed long-term dimming of K giants \citep{Tang2010}, and 
resulted in the discovery of an unusual nova with an outburst (or flare) in 1942 that was followed by a 10 year decline \citep{Tang2012}. Peculiar transients have also been found in the CRTS, e.g. the very long-lasting Type IIn SN 2008iy that took over 400 days to reach its peak brightness \citep{Mahabal2009B}.

Our limiting magnitude is much deeper (Pan-STARRS: $r \sim 23.4$) compared to DASCH ($V \sim$ 15) and we focus specifically on the most extreme events that appeared above, or disappeared below, the detection limit in searches for the most extreme astronomical events and objects. Our timespan is significantly longer than that of the CRTS survey. One may expect to find R Coronae Borealis (R CrB) stars. These are carbon-rich supergiant stars that can dim up to 9 magnitudes with irregular time intervals, where the fading happens on timescales ranging from a few months to years. These eruptive objects have a poorly understood origin, while the most prominent hypothesis is that they formed from mergers of two white dwarfs, or are the result of He flashes in a planetary nebula stars \citep{Clayton2012}. Today we know of $\sim$ 150 R CrB stars in our Galaxy \citep{Tisserand2018} and expect about 5000 to exist.

Highly variable objects such as eclipsing binaries, Cepheids, RR Lyrae, R Coronae Borealis, dwarf novae and highly variable AGN are expected to be detected by VASCO, as their luminosity falls below or raises above the Pan-STARRS detection limit of $r \sim 23.4$. As the limiting magnitude of USNO is around $\sim 20 - 21$, this corresponds to a change of at least 2 magnitudes during the time period of 70 years. Mira variables may vary up to 10 magnitudes on time scales of a few years. Objects similar to these variables may eventually be redisovered during follow-up observations 
with larger telescopes or by patiently waiting for the object to reappear a few months or many years later.

\begin{figure*}
 \centering
   \includegraphics[scale=.65]{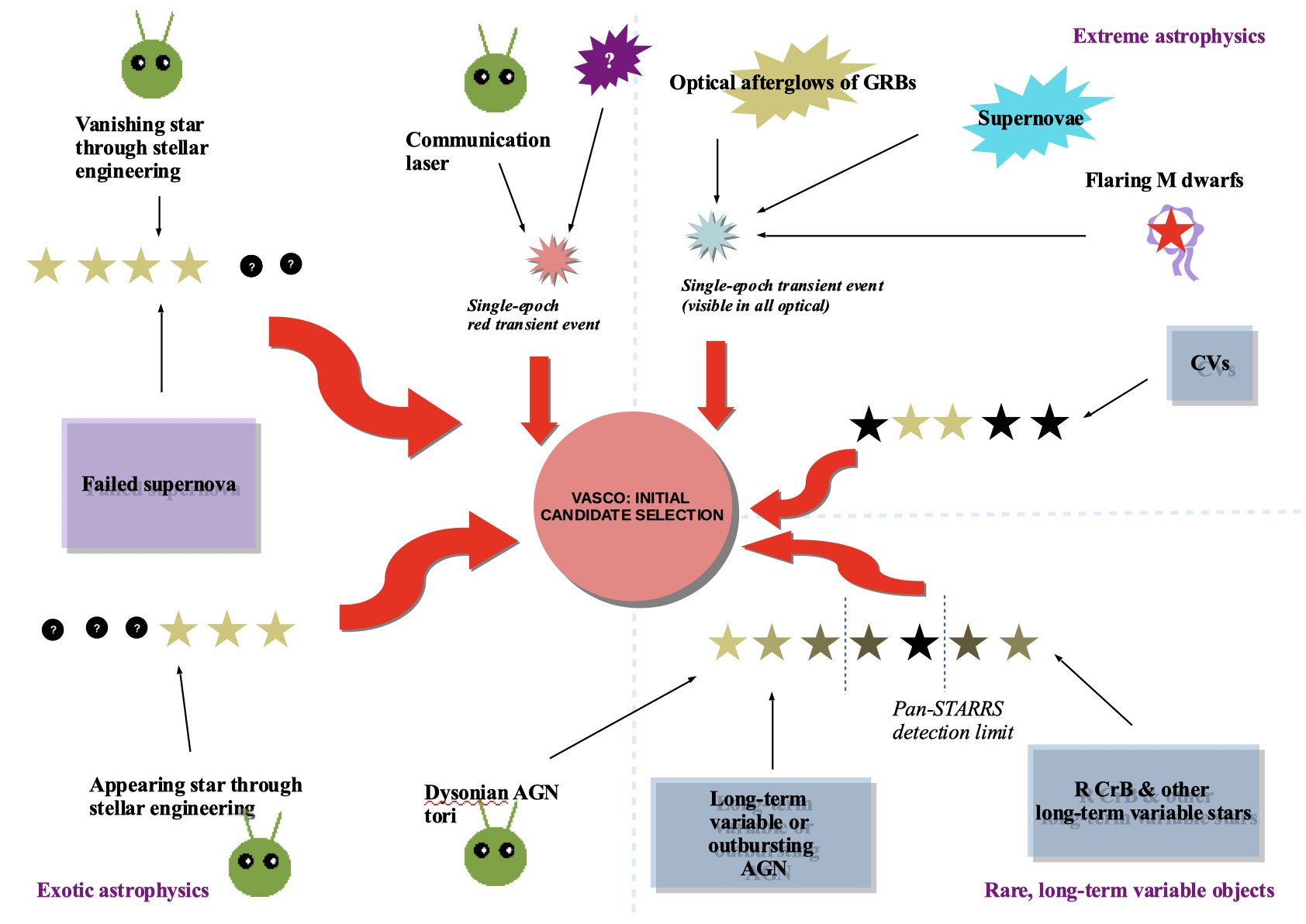}
   \caption{VASCO candidate selection. Once instrumental flaws and errors are removed, we expect different types of objects to be included in the VASCO ``mismatch'' sample. A particular focus is given to USNO objects that either have several detections before vanishing, or to objects that are brighter than $<$ 18.4 magnitudes in USNO and thus have dimmed at least 5 magnitudes.} Rare, long-term variable objects may seem to appear or disappear in the USNO and Pan-STARRS catalogues as they rise above or fall below the detection limit. Among the daily but extreme astrophysical phenomena, we may detect some fast transients only seen at one epoch. Fast transients only seen in the red image could be the result of strongly redshifted transients, less well-known physical phenomena, and also as a result of interstellar communication with red, monochromatic lasers. The VASCO time baseline that probes variability over several decades provides opportunities to study multiple phenomena.
    \label{demoVASCO}
     \end{figure*}

Moreover, VASCO may also discover objects that are only visible in one epoch and then disappear in later surveys. Nearby stars with high proper motion will fall into this category. Outbursts in active galactic nuclei caused by relativistic jet activity or major increases accretion will also give short-term signatures in the optical that fade away in a few months or years e.g. \cite{Prieto1997} or \cite{Mack2009}. Also, transients such as supernovae and tidal disruption events can be detected this way. But natural astrophysical sources are not the only possible sources to discover. Modern Searches for Extraterrestrial Intelligence (SETI) programs are nowadays preparing and executing searches for interstellar optical laser communication, especially in the red and infrared. Therefore, it is of great interest to identify any transient that is only visible once, provided that we can later exclude those events that may be a result of plate defects, cosmic rays and other detection flaws. Figure \ref{demoVASCO} shows what kind of objects we may collect in our first candidate selection. In order to pinpoint the nature of each candidate, one must reconstruct its light curve, which can be done with help of old and modern archives, and by making deeper observations.

%%However, new SETI programs such as the ``Panoramic optical and near-infrared SETI instrument'' (PANOSETI) are nowadays preparing instrumentation to search for short light pulses on timescales of nanoseconds to microseconds that may arise due to interstellar communication \citep{Maire2018,Wright2018}. 

%%Gaia, will have about 70 detections per object in the catalogue, which makes it handy for constructing light curves and to minimize
%%the number of false detections in the older catalogues. 
%%In a worst case scenario, we expect to find 
%%many variable sources near the detection limit of the surveys, as well as a collection of catalogue 
%%mismatches due to false detections in the older catalogues or poor astrometry which will be valuable
%%for improving the existing catalogues.

%%In this first paper, we compare the USNO (Monet et al. 20XX) catalogue that has observations of 1 billion of objects, 
%to Panoramic Survey Telescope and Rapid Response System (PanStarrs) that has detections of 2 billion of objects. 
%5USNO has detections going back to the early 1950s. The PanStarrs, has on the other hand been operating since 2010, 

In this paper, we start by examining the tentative candidate reported by \cite{Villarroel2016}. We present the results of in-depth archival searches, and also some new observations of this object. After examining the candidate, we cross-match the USNO and Pan-STARRS surveys. The current USNO sample is increased by a factor of 60 in comparison to the sample used by \cite{Villarroel2016}, as we use about 60 percent of the USNO catalogue for the cross-matching. In contrast to the previous work, we also include objects with non-zero proper motion. In the Section 3 we discuss the properties of the ``Mismatch Sample''.  We conduct a preliminary analysis of the images in the `SDSS subsample', which includes about 15 percent of the ``Mismatch Sample''. The preliminary list of candidates that resulted from visual examination has been studied at seven epochs (five POSS surveys, SDSS and Pan-STARRS). While this endeavor may include many objects similar to what time-dependent surveys like the Catalina Real-time Sky Transient Survey (CRTS) and Zwicky Transient Facility (ZTF) already detect, we particularly emphasize single-time transients with large amplitudes $\Delta m > $ 5 magnitudes and objects that have been observed in more than one image prior to ``disappearance'' in order to collect the most exotic and extreme phenomena. Finally, we detail the general design and methodology of the VASCO project, as it is currently planned to be carried out over the coming years, including a citizen science project. 

In a separate paper \citep{Pelckmans} we propose a machine-learning based tool aimed to facilitate the planned citizen science project.

\section{The ``vanishing'' star in Villarroel+  2016}\label{sec:illustrative}

%%\subsection{Reexamining a ``vanished star'' found by Villarroel+2016}

Villarroel et al. (2016) identified a candidate, but the candidate was not robust 
enough to make a convincing case for an example of a vanishing star. In the USNO catalogue this object was 
listed as having two detections: one was clearly visible and point-like in the POSS-1 
red band image and the other detection was less clearly visible in the POSS-2 red band. We decided to reexamine it, both reassessing the old observations and by following up with some new imaging obtained with larger telescopes.

\subsection{Observations with CAMELOT at IAC80}

We observed with the IAC80 telescope, which is a part of the Teide Observatory and belongs to the  Instituto de Astrofisica de Canarias (IAC), located at Tenerife Island (Spain). We used the CAMELOT (CAmara MEjorada LIgera del Observatorio del Teide'') instrument in service mode and obtained 9 exposures of 30 minutes each in the red filter. The pixel size is 0.304'' and the limiting magnitude about $\sim$ 24.7 in the Sloan $r$-band.

\subsection{Observations with ALFOSC at NOT}

We made even deeper observations (down to $r \sim 25.5 - 26$) with the help of the Alhambra Faint Object Spectrograph and Camera (ALFOSC) instrument at the Nordic Optical Telescope (NOT, La Palma, Spain) in service mode and fast-track observations. The goal
was to carry out deep enough observations to be able to detect a point source at the 25th magnitude level with a signal-to-noise ratio of at least 9 or 10, using deep Gunn r'-band imaging. Assuming an airmass of 1.5, seeing of 1 arcsec and a grey night, we estimated that about four hours of observation time were needed. Six exposures of 900 seconds were taken. For the resulting images, the pixel size was 0.214" and the limiting magnitude about $\sim$ 25.5 - 26.0 in the $r$-filter.

\subsection{Results from the observations}

 \begin{figure*}
 \centering
   \includegraphics[scale=0.2]{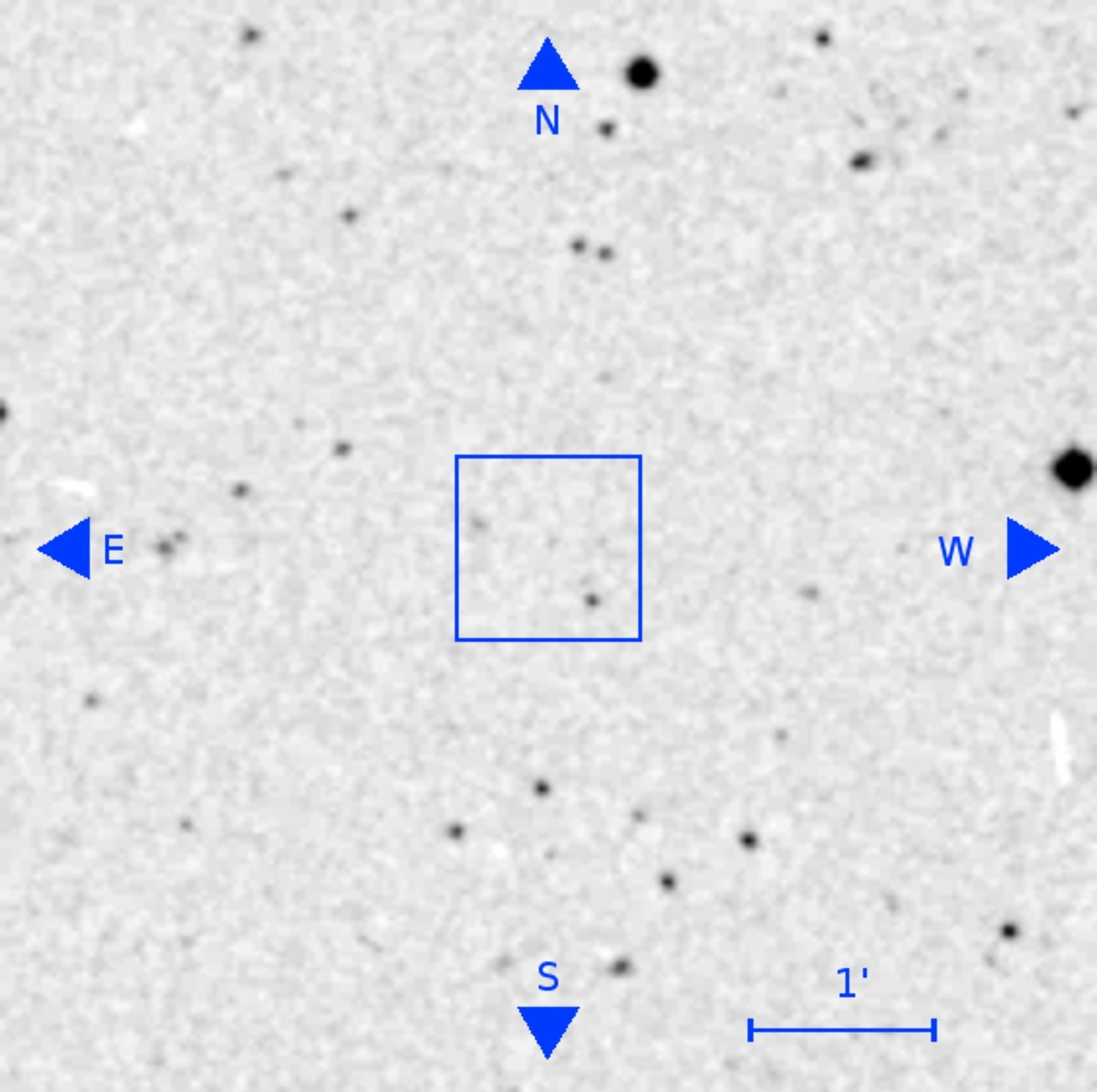}
   \caption{The POSS-I E red observation of the Villarroel+ 2016 object, which is 
   centered in the small square.  The object in the lower right quadrant is another,
    brighter field star.}
    \label{Oldimage}
     \end{figure*}
     
\begin{figure*}
 \centering
   \includegraphics[scale=.20]{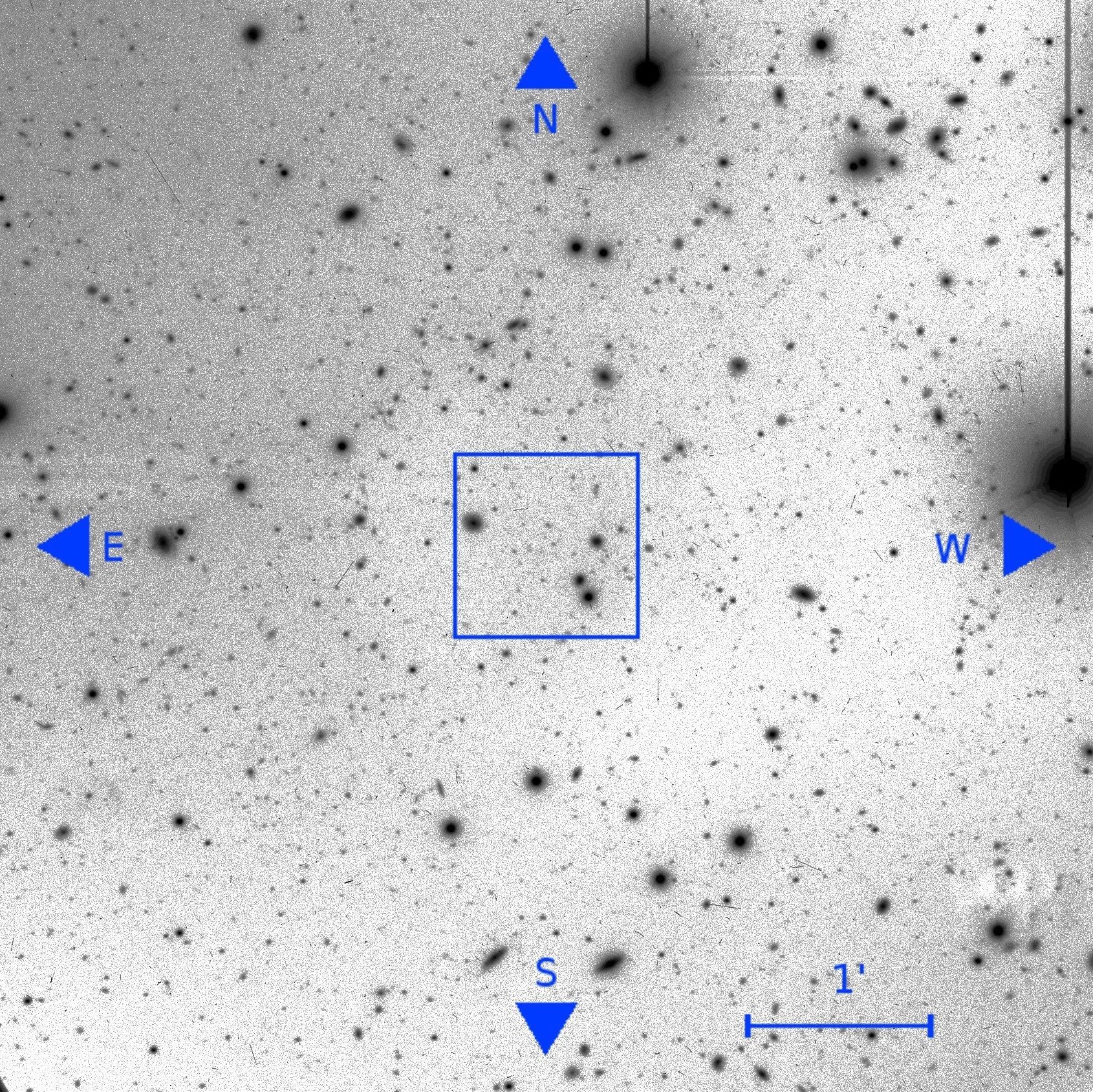}
   \caption{The NOT observation of the Villarroel+ 2016 object. The image was constructed by summing two observing blocks taken 16 May 2018 and 23 May 2018. There are two potential counterparts 1.4 arcsec northwest and 2.4 arcsec southwest of the USNO object's original position.}
    \label{NOTimage}
     \end{figure*}
     
We summarize what we know about the existing observations of the object so far. See Table \ref{OldObservations}. The table presents old archival observations as well as new ones we have performed during the follow up.

We first examine the old POSS images. As we can see from the table, the minimum requirement of two detections (on which USNO is based) is not clear for this particular object. Only one strong confirmation (POSS-I E plate) exists. Unlike an artifact, the object appears to be point-source-like in the POSS-I E plate. See Figure \ref{Oldimage}. One possibility is that this object is a star with significant proper motion and moved entirely out of the image.
     
We compare the POSS-1 E image with the new images taken with the NOT. See Figure \ref{NOTimage}. In the NOT imagery we find two objects very close to the original USNO location. One of the objects is located 2.4 arcsec southwest of the USNO object, and the second is 1.4 arcsec northwest of the USNO object. However, the resolution in the POSS-I E band is about 1.7 arcsec per pixel, and the displacement of the two reported objects is therefore within the error, in particular for the closer object only 1.4 arcsec away.

The colors may give a clue. The original USNO object was only seen in the red band. While the nearby WISE counterpart is seen both in the blue band with the Magellan telescope and in the red band with NOT, the NOT objects can only be seen in the red. This may support the hypothesis that the object from the 1950s and one of the objects seen in NOT are likely to be the same. But if so, the brighter (southwest) object has dropped about 4.2 to 4.3 magnitudes in the $r$-band and also moved a bit. 

%%From the limiting magnitude in the blue from USNO ($b \sim 21$) and the measured magnitude we know that the lower limit of color was about $b - v > 1.3$ in the

\begin{table*}[ht]
\caption{Summary of the observations. {Possible detections, detection limits\footnote{Limiting magnitudes for POSS are taken from \citep{DPOSS}.} and date of observations are reported.} We use the DSS Plate Finder to retrieve the images used in the USNO database. For duplicate images we report the longest exposure time. As can be seen from the images, the object in the POSS-I E red image is point-source like. For the same object we find that the positional error of the object is 5.5 arcsec. This means the WISE counterpart is a possible counterpart. However, in the significantly deeper NOT images we find both the WISE counterpart and two possible candidates very close to the position of the original USNO object.}
\centering
\begin{tabular}{c c c c}
\hline\hline
\multicolumn{4}{c}{Samples} \\
\hline\hline
Archive/telescope & Detection? & Detection limit [mag] & Date of observation\\
POSS-I O blue $\sim$ 4100 \AA  (Palomar) & no & 21 & 1950-03-16 \\[0.1ex]
POSS-I E red $\sim$ 6500 \AA  (Palomar) & yes, 19.7 mag & 20 & 1950-03-16\\[0.1ex]
POSS-II J blue $\sim$ 4400 \AA  (Palomar) & no & 22.5 & 1993-05-19\\[0.1ex]
POSS-II F red (Palomar) $\sim$ 6600 \AA  (Palomar) & offset/dirt? & 20.8 & 1993-03-22\\[0.1ex]
POSS-II N IR $\sim$ 8400 \AA   (Palomar) & faint detection/noise? & 19.5 & 1993-03-04\\[0.1ex]
Quick-V Northern & faint detection/noise? & ? & 1982-05-19\\[0.1ex]
SDSS (New Mexico 2.5 m) & no & 23 & ?\\[0.1ex]
Pan-STARRS & no & 23.4 & ?\\[0.1ex]
Gaia & no & 21 & ?\\[0.1ex]
GALEX-5 & no & ? & ?\\[0.1ex]
Ukraine VO\footnote{Ukraine VO archives are described by \cite{Vavilova2012} and \cite{Vavilova2016}.}\footnote{http://gua.db.ukr-vo.org/archivespecial.php}\footnote{http://ukr-vo.org/digarchives/index.php?b5\&1} & no & $\sim$ 17 & 1982-05-26\\[0.1ex]
Ukraine VO & no & $\sim$ 17 & 1987-04-23\\[0.1ex]
Ukraine VO & no & $\sim$ 17 & 1991-05-11\\[0.1ex]
Ukraine VO & no & $\sim$ 17 & 1993-05-10\\[0.1ex]
Ukraine VO & no & $\sim$ 17 & 1993-05-11\\[0.1ex]
WISE & Counterpart 5.8 arcsec north? & ? & ?\\[0.1ex]
%%41 cm Ritchey-Cretien (UAA), red & no & ? & ?\\[0.1ex]
%%Westerlund telescope (90cm), red & something at noise level? & ? & 2017-03-18 \\[0.1ex]
CAMELOT (La Palma, 1.0m), red & something at noise level? & 24.7 & 2018-05-09\\[0.1ex]
ALFOSC (NOT, 2.5m), red, obs. block 1 & 24.26 $\pm$ 0.02 (southern object) & $\sim$ 25.5-26 & 2018-05-16\\[0.1ex]
ALFOSC (NOT, 2.5m), red, obs. block 2 & 24.26 $\pm$ 0.02 (southern object) & $\sim$ 25.5-26 & 2018-05-23\\[0.1ex]
ALFOSC (NOT, 2.5m), red, obs. block 1 & 25.18 $\pm$ 0.03 (northern object) & $\sim$ 25.5-26 & 2018-05-16\\[0.1ex]
ALFOSC (NOT, 2.5m), red, obs. block 2 & 25.18 $\pm$ 0.03 (northern object) & $\sim$ 25.5-26 & 2018-05-23\\[0.1ex]
Magellan telescope (Baade 6.5m), blue & no & ? & 2018-06-05\\[0.1ex]
\hline
\end{tabular}
\label{OldObservations}
\end{table*}

One may wonder what is the probability of observing a new, unrelated object with NOT 
within 2.5 arcseconds of the stated position in USNO, if going 4.2 magnitudes deeper. However, probability estimates of this sort are of little help when we search deliberately for outliers in big datasets covering billions of objects.

%%Which of the two objects in the NOT image that correspond to the ``vanished star'' -- and if any at all-- remains unsolved. In order to know if any of the NOT objects is truly the same object or not, we would need to measure the proper motion of two closest candidates (within 2 arcsec from the USNO object) by doing repeated, careful photometry the coming 10 years and see if the movement 
%%of any is consistent with the displacement, regarding the direction and size of the shift.

%%A positional shift could also happen, 
%%if the POSS-1 E observation was actually a supernova that happened in the outskirts of a (now faint) galaxy.

\subsection{Could it have moved?}

One possibility is that our target is a star with a fairly high proper motion (despite being catalogued in USNO as having no proper motion), and that it has moved substantially from its original position. If we compare the two different red plates from the POSS within a reasonable angular distance, we should be able to find the missing object by seeing an appearing object in the later POSS image from 1993.

Assuming a maximum proper motion of 6 arcseconds per year (slightly larger offset than the positional error of 5.5 arcsec), we know that between 1950 and 1993 the star will not have moved more than 4.3 arcminutes in any direction during these years. We therefore extract red filter images from POSS-1 (from 1950) and POSS-2 (1993) with a field of view corresponding to 9 x 9 arcminutes. We inspect these visually by ``blinking'' them. The few objects that ``appear'' in the later epoch turn out to also exist in the SDSS images, which means they simply were not resolved in the previous epoch. No
other ``appearing'' objects could be seen in the later epoch, which means we can quite safely reject the hypothesis of a fast moving star. Solar system objects typically are bluer (as they shine by reflected sunlight), although exceptions of course exist. From the DSS Plate Finder \footnote{https://archive.stsci.edu/cgi-bin/dss\_plate\_finder} we see that the POSS-1 E red and POSS-1 O blue image were taken with a time difference of about half an hour, but nothing is visible in the blue image at the position of the star. Given the exposure time of 45 minutes of the POSS-1 E red image, if our object were an asteroid that quickly moved out of the field, it would have left a stripe (and not be point-like).

\subsection{Was it possibly an image defect?}\label{sec:defects}

The final hypothesis that could rule out the idea of a transient or variable event in the 1950s plate is the simplest explanation of all: plate defects in the old photographic plates from POSS-1. While the USNO-B1.0 should be cleaned up from a fair number of these artifacts, and a separate list by \cite{Barron2008} could have included our target but did not, our target still has survived thanks to the two detections listed in USNO. 
%Also, we note that plate defects are very seldom point-like.

We reanalyze the POSS images based on high-resolution data from STScI Digitized Sky Survey\footnote{https://archive.stsci.edu/cgi-bin/dss\_plate\_finder}.
We see several things: only the detection from the POSS-1 E red plate taken on 16 March 1950 has a 
secure detection. The second detection -- which we believe is based on the POSS-II F image from 22
March 1993 -- is slightly offset and possibly not the same object (even if listed as the same; here the low resolution may have played a role). Of all the other images available on that server covering that particular sky region -- Quick-V Northern (1982), Poss-I O (blue, 1950), POSS-II Blue (1986), POSS-II N (1993), POSS-II N (1996) -- none of them convincingly shows the object. Some hints of an object may be seen at the given position in the Quick-V Northern image from 1982, but not in a way that would allow us to confirm the detection quantitatively as the signal-to-noise ratio is very low. 

While plate defects in USNO very seldom are star-like \citep{Gaensler}, some of the star-like sources could in principle be photographic plate defects. 

 These defects can be created when a small dust particle sticks to the plate during the exposure, or when microspots form after years of storage. \cite{Greiner} proposed examining original plates with a microscope in reflected light to help sort out which events we see are real astronomical events, and which ones that are pure plate defects. The best way to be sure when dealing with old photographic plate material is to investigate the photographic plates themselves under a microscope.

Unfortunately, we do not have access to the original plates. However, by comparing the point spread function (PSF) of the object to the PSF of typical stars in the same field, one can see if the object is likely to be a plate flaw or a real star. See Section \ref{sec:inspection}. If an object has a considerably smaller PSF than a real star as measured on the given photographic plate, it may be discarded as a plate flaw. Our object appears to be like many astronomical point sources and has a PSF comparable to the real stars on the plate. This suggests that it is not a plate defect.

\section{New searches, new samples}\label{sec:methods}

%%In this section we reexamine the ``vanished star'' found by Villarroel+ 2016. In the USNO catalogue this object was listed as having two detections, one which was clearly visible and point-like in the POSS-1 red band image, and another detection which was less clearly visible in the POSS-2 red band. 
We continue our exploratory journey by cross-matching the USNO and Pan-STARRS catalogues in searches for better candidates.

\subsection{Starting samples}\label{sec:new}

The USNO-B1.0 catalogue \citep{Monet2003} presents the best old sky survey we can use in the optical as it goes deep enough (r $\sim$ 20) and contains 1 billion astronomical objects. It has all-sky coverage. This allows us to find astronomical transients that occurred before the birth of the all-sky transient surveys. Each object is \textit{supposedly} detected at least twice in two widely separated epochs in the Palomar Sky Survey (POSS).\footnote{POSS1 was carried out from 1950 to 1966, and POSS2 from 1987} to 1999. The data were obtained in one blue band and one red band, and for some objects, also in the infrared.
The Pan-STARRS catalogue has about 2 to 3 billion objects and is at present the largest digital sky survey with observations started in 2010. It covers the entire sky down to declinations dec$\sim -30$ deg.  Information about the Pan-STARRS data products is described in a series of articles  \citep{Chambers,MagnierA,MagnierB,MagnierC,Waters2016,Flewelling2016}. Our PS1 dataset is an offline version kindly provided by the Pan-STARRS collaboration.

One may wonder if it would not be better to directly search for vanishing or appearing objects only using internally consistent datasets like Gaia or Pan-STARRS, containing about 2 to 3 billion objects each, where each object has photometry done multiple times during 5 years of observations. The CRTS has homogeneous data and time baselines up to 14 years with CCD data. However, when one compares the digitized sky from USNO plates with the sky from Pan-STARRS, the significantly longer time span ($\sim$ 70 years) changes the effective volume of the dataset over which any event could have been observed. Extremely rare events are much more likely to be found in surveys that combine both a long time baseline and deep photometry. The DASCH survey may have 100 years of photometry, but has a limiting magnitude around $V \sim$ 15. The longer time span allows us to discover extreme variables with a characteristic timescale of several decades, longer than the typical five years of iPTF. An example of a vanishing-star event that is not expected to happen in the Milky Way more often than once every few hundred years, is the hypothetical failed-supernova event (see Appendix). The VASCO time baseline and depth in photometry makes it possible to discover such events. Of course, we expect also to detect many objects that vary on shorter time scales.

\subsection{Cross-matching the USNO and Pan-STARRS catalogues}\label{sec:crossmatch}

The goal of the cross-matching algorithm used in this paper is to make a list of USNO objects that do not have a Pan-STARRS counterpart within a certain distance threshold (e.g. 30 arcseconds). In the \citep{Villarroel2016} paper the size of the starting samples used were, on average, about $\sim$ 10 million USNO objects. However, using the full catalogues of USNO-B1.0 and Pan-STARRS DR1 is more of a practical challenge, as the databases we have make up about 1 TB in size (roughly 300 GB and 700 GB each, respectively). This creates a problem of 
efficiency in the cross-matching process, which could last unacceptably long if not done smartly.

We use a 3 TB cloud environment provided by the Uppsala Multidisciplinary Center for 
Advanced Computational Science (UPPMAX), which is part of the Swedish National 
Infrastructure for Computing (SNIC). The cross-matching is done in the environment of SQlite3, and carried out by parallellizing the cross-matching process by breaking down the USNO and Pan-STARRS databases 
into many smaller ones with the help of smart index methods. This enables the cross-matching process to be done effectively in smaller subsets compared to using the whole databases. All the technical details of the cross-matching are described by \cite{Soodla2019}.

 \begin{figure*}
   \centering
  \includegraphics[scale=0.5]{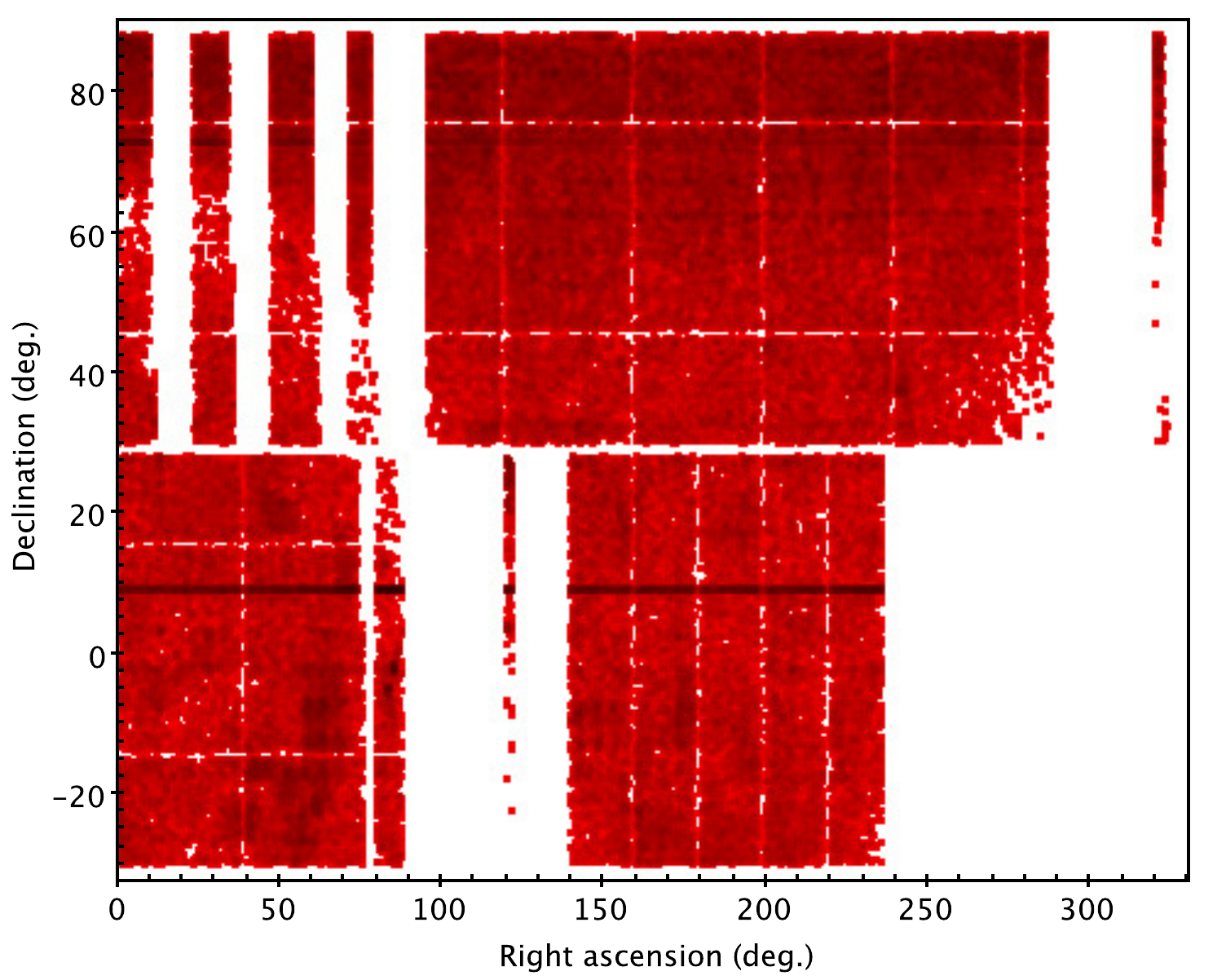}
  \includegraphics[scale=0.5]{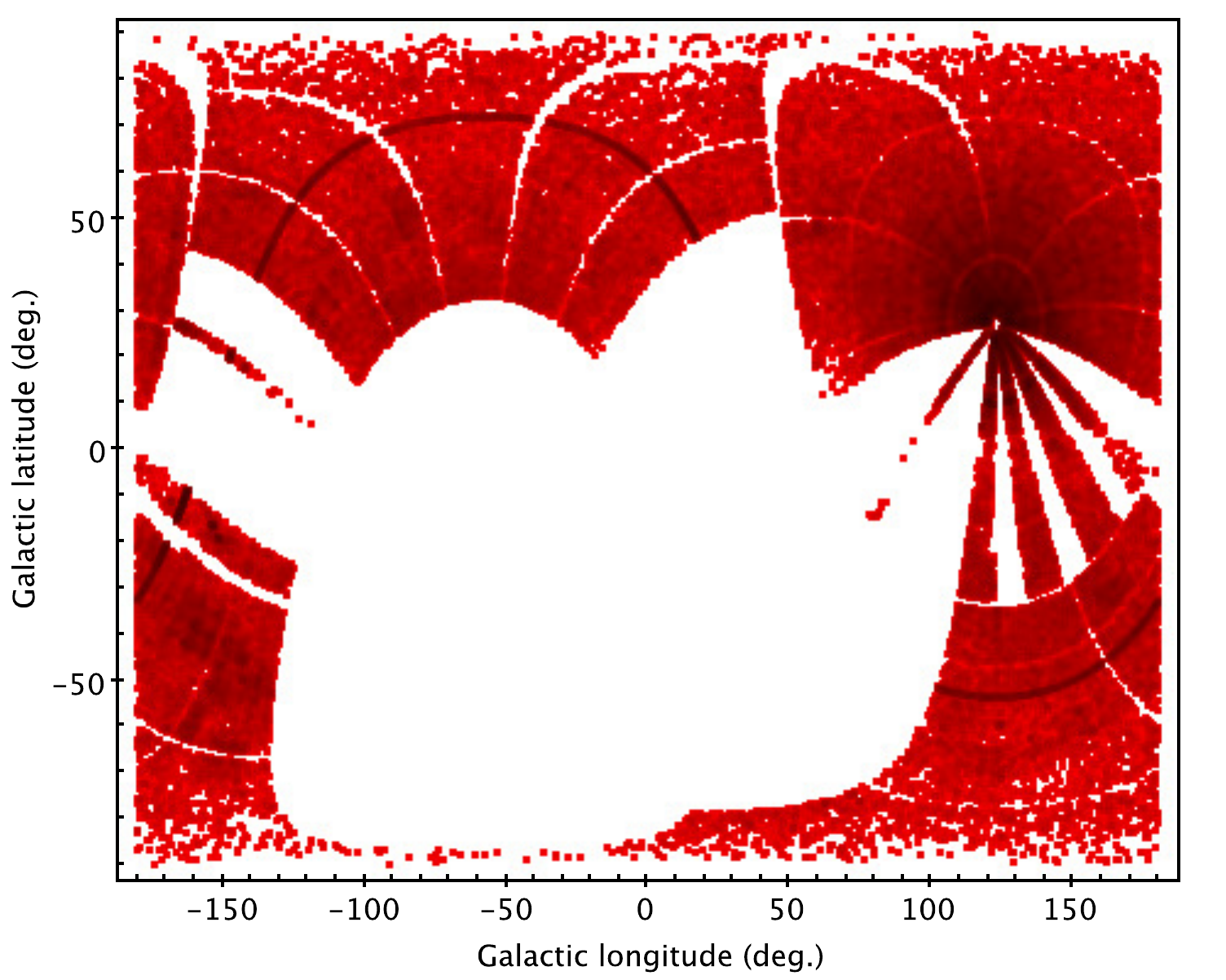}
  \caption{\label{distribution1} The distribution of the $\sim$ 150,000 candidates on the sky. The left plot shows equatorial coordinates, and the right plot shows Galactic longitudes. Some regions are unsampled in the cross-match, as can be seen by the empty regions, in particular ra(240,360) and dec(-30,30) in equatorial coordinates. 
  }
   \end{figure*}

The cross-matching procedure in VASCO differs from a traditional cross-matching between two catalogues, as we are searching for missing objects rather than corresponding objects.

In a traditional cross-match, one uses an object from catalogue A and tries to identify the same object in catalogue B using the coordinates (and additional properties like fluxes, surface densities of sources, etc). Due to proper motion, variability and many other factors, it can be quite challenging to verify if the object within a certain radius in catalogue B is the same object. A typical cross-match radius in traditional projects is 3 to 5 arc seconds. 
If one used instead a large cross-matching radius (e.g. 30 arcsec), there are often several possible matches, which means we may have a number of \textit{false positives} among the cross-matches, and we have included spurious objects in the resulting catalogue.

In our particular case, the cross-matching is not a traditional cross-match. When we take an object from catalogue A and try to look for a ``vanishing'' object in catalogue B, we only care to know that no object at all resides at the given position in the second catalogue B. If one uses a small ``cross-match'' radius like 5 arc seconds, this leads to a large number of mismatches as various astrometric issues enter, including the proper motion of objects. 
However, by extending the ``cross-match'' radius to 30 arc seconds,
one implicitly takes care of proper-motion related issues, except possibly for nearby red dwarfs or white dwarfs.
That would make sure that USNO objects with proper motions less than $< $0.4 arcsec/yr over a 70 yr baseline are directly excluded from the resulting ``mismatch'' sample. The downside with this method is that one misses out on potential mismatches as \textit{false negatives} enter the picture.  This means that with a large cross-match radius our mismatches are very likely to be real mismatches, but we underestimate the number of mismatches (and hence we miss candidates). For objects with proper motions larger than $\sim $0.4 arcsec/yr, the displacement in coordinates is visible and easy to identify by blinking images. See Section \ref{sec:propermotion}.
%%We just make sure that there is not a single object in Pan-STARRS exists within this radius.}
   
From the USNO and Pan-STARRS J2000 coordinates we determine whether a counterpart exists or not within a certain angular distance. The USNO objects not having a counterpart we list as ``mismatches'' together with the closest Pan-STARRS neighbor. Only the positional proximity is used. In this early study we covered only 60 percent of the sky (about 600 million USNO objects) due to limitations in computing time, and some regions are left out in the cross-match, as seen in Figure \ref{distribution1}.

Using a 30 arcsecond threshold (a limit set by the available computing time in the cloud environment), we find 426,975 mismatches (corresponding to a mismatch rate of 0.074 percent). Correcting for differences in sky coverage between USNO and Pan-STARRS by removing all objects with declinations $<- 30$ degrees, 151,193 of the mismatches can be considered for further investigation. The mismatch rate is within the range of various data processing artifacts existing in sky surveys, and among these artifacts we must search for real candidates.

\subsubsection{Treatment of high proper motion objects}\label{sec:propermotion}

\begin{figure*}
 \centering
   \includegraphics[scale=0.5]{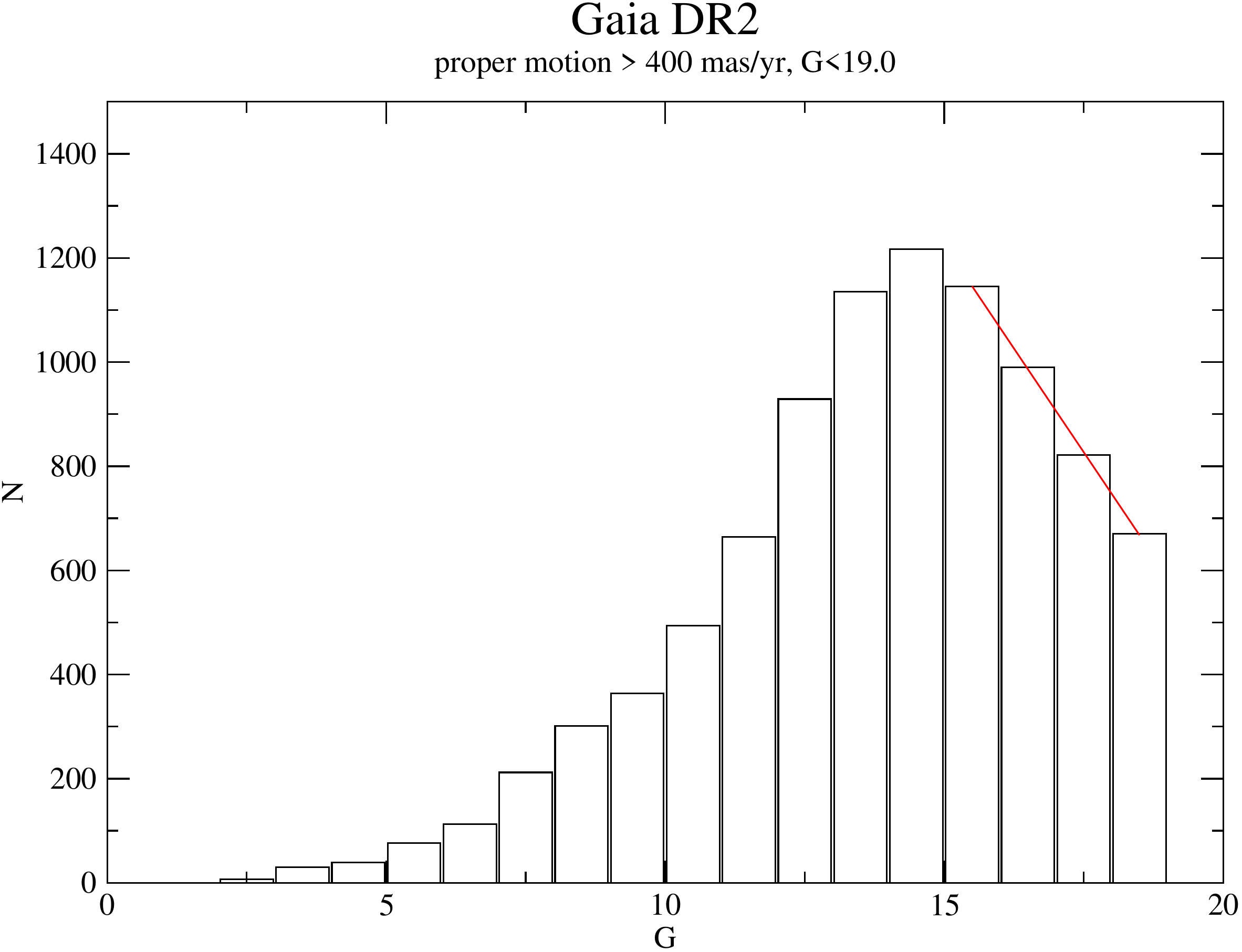}
     \caption{Proper motions in Gaia. We show all objects (9208) with high proper motions larger than 400 m.a.s. per year. Only 2482 are in the magnitude range $16 < G < 19$.}
     \label{Gaiapropermotions}
     \end{figure*}

For the 151,193 mismatches we first must ask: how many of these are only the result
of a star just moving away over the last 70 years? We approach the problem by estimating the number of objects that would escape our 30 arcsec cross-match radius. See Section \ref{sec:crossmatch}. A 30 arcsec cross-match radius over a 70 year timeline translates to proper motions larger than 0.4 arcsec per year. We therefore use the Gaia Data Release 2 (DR2) catalogue \citep{Gaia2016,Gaia2018} to obtain all catalogue objects that have $\mu_{tot} >$ 0.4 arcsec per year. The catalogue is complete down to $g \sim 19$. We plot a histogram of their magnitudes in Fig. \ref{Gaiapropermotions}.

As we later find that 95 percent of the objects in our mismatch sample are fainter than mag 16 (see Figure \ref{Magnitudes}), we estimate the number of objects with $g > 16$ in Gaia DR2. Between $16 < g < 19$ there are 2,482 objects. We extrapolate that the number of objects in the last bin $19 < g < 20$ is $\sim$ 500, which means that the number of objects in Gaia DR2 with  $g > 16$ and proper motions larger than 0.4 arcsec per year is roughly $\sim$ 3,000. Correcting for the sky coverage used in our cross-match, this decreases the number by a factor of two, meaning that we may expect around $\sim 1,500$ objects with high proper motion to contaminate our 150,000 mismatches. These objects can, however, be spotted when comparing the images and their surrounding fields. 

We filter the mismatches by one additional proper motion limit. This limit is set by the image field we are prepared to investigate visually later on. Any USNO star that moves away will move at a limited angle distance per unit time, and will be seen as an ``appearing'' case in the corresponding Pan-STARRS survey at a different location, likely with the same colors and magnitude (unless also variable). 

The Gaia survey has shown that there are only 9 stars known to us with proper motions larger than 5 arcsec per year. Over 70 years of time this corresponds to a movement of about 7 arc minutes between the POSS-1 and Pan-STARRS images.
For a radius of 7 arc minutes it would therefore be wise to use image fields of size 15 arc minutes when we compare the images, if we want to keep all objects with proper motions up to 5 arcsec per year. We note that USNO's proper motions carry much larger uncertainties than those of Gaia, and removing all objects with proper motions in USNO larger than 5 arcsec per year (about 130 objects in the mismatch sample) would leave us 151,063 objects.

For practical purposes, we shall use 5 x 5 arc minute images (a search radius of 2.5 arcmin). We filter the data so that we restrict the listed USNO proper motions to be less than 4.3 arc seconds per year, leaving 151,038 objects in our mismatch sample.

\subsection{Visually inspecting a subset with the SDSS}\label{sec:visual}

One of the ways to investigate the 151,038 mismatches is to look at those missing in the Sloan Digital Sky Survey (SDSS) Data Release 12 (DR12). The SDSS only covers the Northern Hemisphere, and therefore approximately half of the objects have not been observed in both surveys. Also, the Sloan Digital Sky Survey started at an earlier epoch than Pan-STARRS, which reduces the time window for a potential disappearance
 by about 10 years. Consequently, vanishing events that have happened in the last decade may remain undetected. 

In order to cross-match with the SDSS DR12, we use the CasJobs interface\footnote{https://skyserver.sdss.org/CasJobs/},
upload our coordinates to the server and use the Footprint function to check if a coordinate is within the SDSS scanned field. We see that 64,475 objects out of 151,038 can be found within the scanned field of the SDSS. These objects we re-upload to the CasJobs, and then do a closest neighbor search with a radius of 0.08 arcmin (5 arcsec). About 23,667 objects have no detectable closest counterpart in this search zone\footnote{Using the fGetNearestObjEq(m.ra,m.dec,0.08)}. This means, that roughly one third of our candidates remain using 5 arcsec as a cross-match radius.

Here, we carry out a similar analysis as in Section \ref{sec:propermotion}, taking into account that SDSS only covers $\sim$ one-fourth of the sky, and see that the estimated number of expected mismatches for proper motions necessary to exceed a 5 arcsec cross-match ($\mu_{tot} > 0.080$ arcsec per year) is large -- about 125,000 objects. However, most of these objects will not be a part of our visual subset. That previously used a 30 arcsec cross-match radius with Pan-STARRS.

At this point it would have been useful to employ image differencing software to compare the images to identify any obvious differences in pairs of very similar images. But, as we compare images made with widely different telescopes, instrumentation and methods (photographic vs CCD), we do not gain much advantage by doing this step. Moreover, the hard drive space required to download the many fits files is prohibitive; one thousand images occupy $\sim$ 1 TB. Therefore, we have inspected each of 23,667 candidates
individually by visually comparing the images found in DSS1 \footnote{http://catserver.ing.iac.es/dss1/}, 
STSCI archive\footnote{https://archive.stsci.edu} and SDSS Explorer\footnote{http://casjobs.sdss.org/dr15/en/tools/explore/Summary.aspx?}. 
First, we used the SDSS Explorer list-view to remove all objects that had an obvious
flaw such as a bright star or dead stripe in the SDSS image. See \cite{Villarroel2016} for details. This left 6359 objects, where no obvious flaw was causing the mismatch. In the next stage we individually examined the 6359 images in the DSS1 and only kept those that had an object in the center of the image, in order to remove false positives among the original USNO objects. This left 1691 candidates that had something clearly visible in the center of the DSS1 image. The SDSS subset effectively covers about 90 million stars from the USNO starting sample.

\section{Results from the new searches}\label{sec:results}

\subsection{The sample properties of the new ``Mismatch Sample''}
One possibility is that the mismatches we have found represent objects with some typical problems. For instance, our objects could have larger average proper motion than reported in USNO. Also, our objects could have fewer detections associated with them, in comparison to the ``average'' USNO object, which leads to a number of false positives. Therefore, we investigate some basic properties of the Mismatch Sample, and compare them to 49,999 typical USNO objects, randomly selected from the entire USNO catalogue.

\begin{figure*}
 \centering
   \includegraphics[scale=0.5]{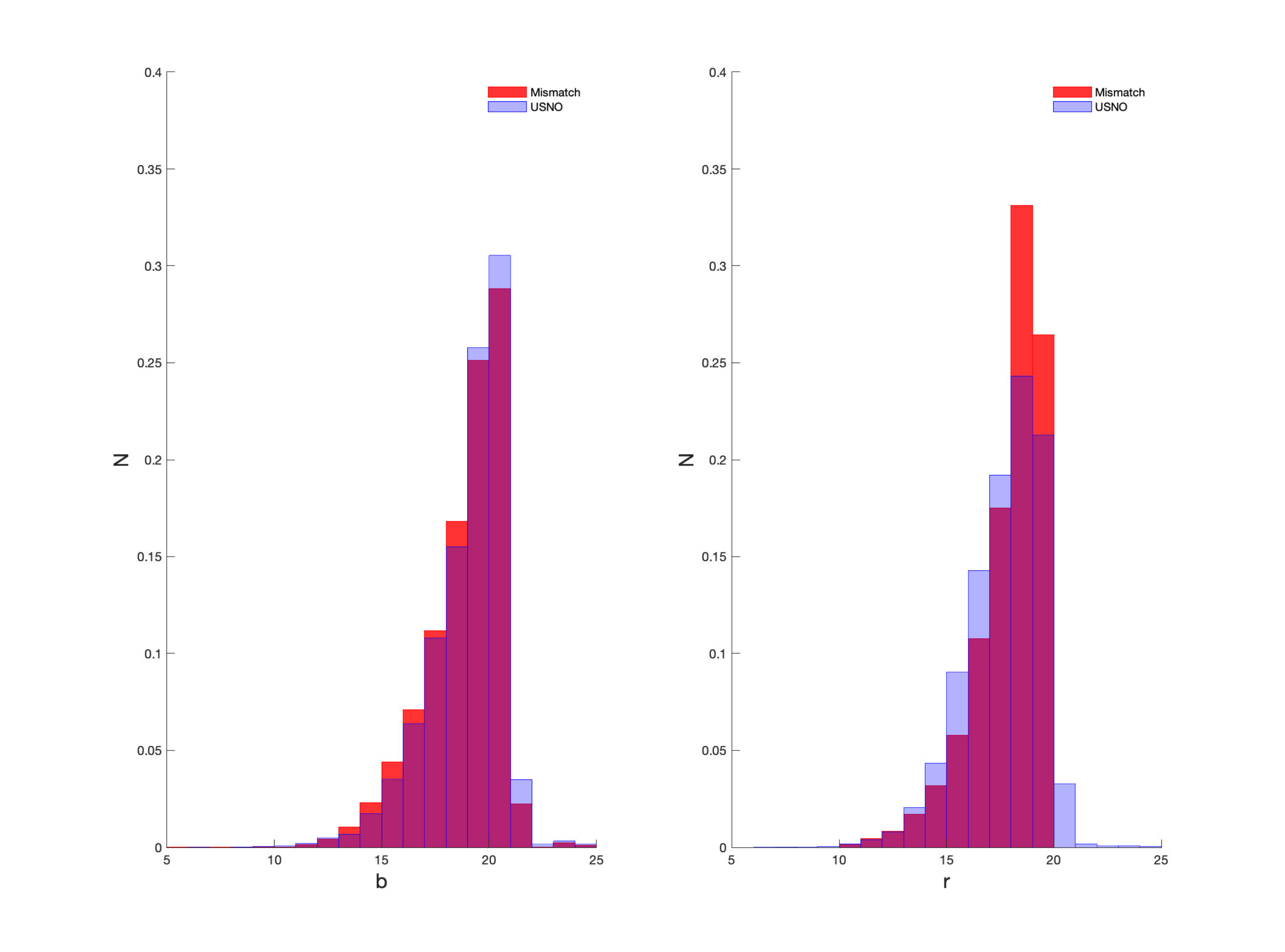}
     \caption{The apparent magnitudes of the objects in two different filters, b (blue) and r (red) filters. The left column shows the magnitudes from the first epoch, and the right column from the second epoch.
     The Mismatch Sample (left) is compared to 50,000 randomly-selected USNO objects (right). The histograms are normalised and zoomed.}
     \label{Magnitudes}
     \end{figure*}

Figure \ref{Magnitudes} shows histograms over the apparent magnitudes (blue and red band) for the Mismatch Sample and the $\sim$50,000 randomly selected USNO objects. We see that the mean value in the blue band is $b \sim$ 18.85 $\pm$ 0.01 (Mismatch Sample) and $b \sim$ 19.01 $\pm$ 0.01 (USNO) in the first epoch (POSS1: years 1949 to 1966). For the red band the average is $r \sim$ 17.86 $\pm$ 0.004 (Mismatch Sample) and $r \sim$ 17.72 $\pm$ 0.01 (USNO). A two-sample Kolmogoroff-Smirnoff test reveals a small but statistically significant difference between the samples, where the mismatch objects are slightly fainter in the red, but brighter in the blue band.

\begin{figure*}
 \centering
   \includegraphics[scale=.5]{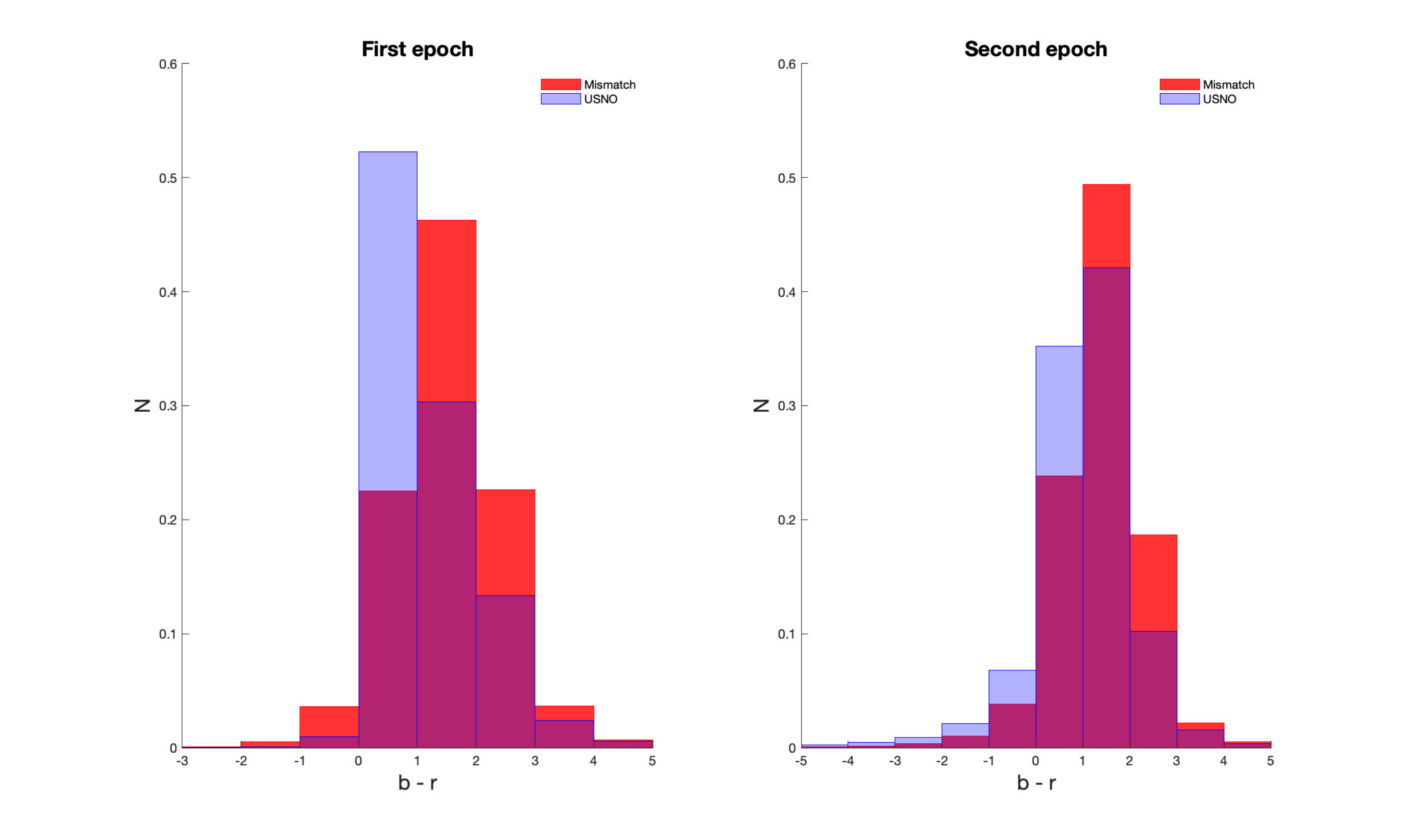}
     \caption{The $b - r$ of the objects in two different filters, b (blue) and r (red) filters. 
     The left column shows the colors from the first epoch, and the right column from the second epoch.
     The Mismatch Sample (left) is compared to 50000 randomly-selected USNO objects (right). The histograms are normalised and zoomed.}
     \label{Color}
     \end{figure*}

As a next step, we consider the colors. See Figure \ref{Color}. {The used filters O (POSS-I blue), E (POSS-I red), J (POSS-II blue) and F (POSS-II red) have the effective wavelengths of 4100 \AA , 6500 \AA, 4700 \AA\ and 6600 \AA .} The samples could possibly come from two different color distributions. Indeed, a two-sample Kolmogoroff-Smirnoff test shows a statistically significant difference if testing with the nominal value of $\alpha < 0.05$. The average colors in the first epoch are $b - r \sim$ 1.49 $\pm$ 0.003 (mismatch) versus $b - r \sim$ 0.94 $\pm$ 0.01 (USNO sample). In the second epoch, the corresponding mean values are is $b - r \sim$ 1.37 $\pm$ 0.003 (mismatch) and $b - r \sim$ 0.99 $\pm$ 0.01. As the colors at the faint end may be uncertain, we also considered the corresponding color indices when using magnitudes brighter than 18 mag. We see that the average color differences in the first epoch are more pronounced for mags $<$ 18, with $b - r \sim$ 1.22 $\pm$ 0.005 (mismatch) versus $b - r \sim$ 0.400 $\pm$ 0.01 (USNO sample).
%%{[Rounded off to 0.001 
%%mag, and proper motion values below are rounded off to 0.001 arcsec/yr.]}

\begin{figure*}
 \centering
   \includegraphics[scale=.6]{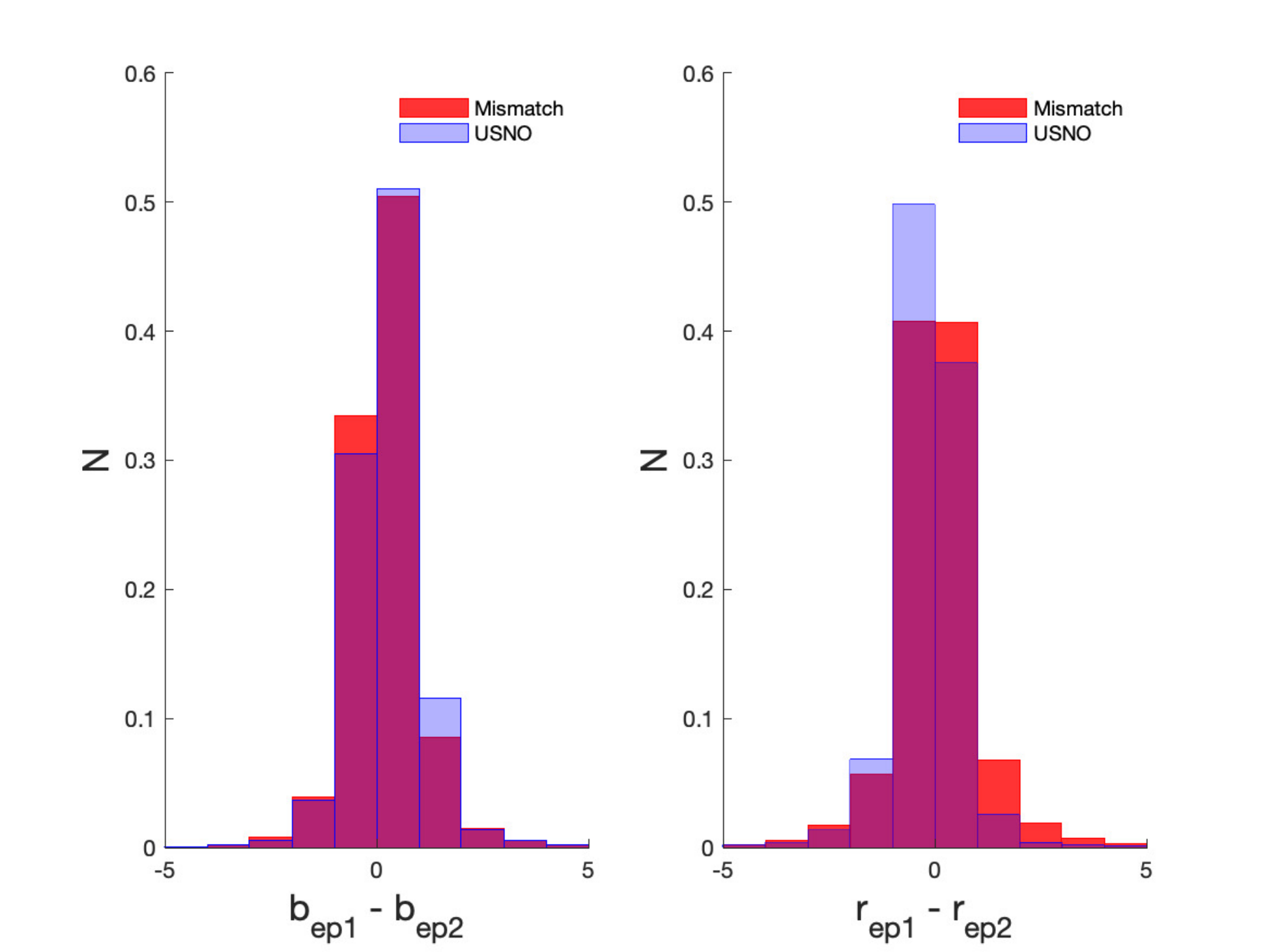}
     \caption{The difference in apparent magnitudes between the first and the second epoch, as seen in two different USNO bands, b (blue) and r (red) filters. The Mismatch Sample (left) is compared to 
     50000 randomly-selected USNO objects. The histograms are normalised and zoomed.}
     \label{Variable}
     \end{figure*}

We also compare variability separately in the two different bands. We show the difference between the brightness in the first and second epochs in Figure \ref{Variable}. In the blue band, the average change in magnitude is 0.19 $\pm$ 0.004 mag (mismatch) or 0.27 $\pm$ 0.007 (USNO). While the difference is significant enough to be noted in a 
 Kolmogoroff-Smirnoff test, it is not particularly large and totally within the instrumental or calibration errors of USNO \citep{Gaensler}. In the red band the difference is: 0.03 $\pm$ 0.003 mag (mismatch) and $-$0.17 $\pm$ 0.005 mag (USNO). The difference here is also statistically significant in a two-sample Kolmogoroff-Smirnoff test. However, they are very likely to be the result of photometric calibration issues in the USNO survey, where the standard deviation of magnitudes in any band is about 0.3 mag but the systematic errors can be up to several magnitudes \citep{Monet2003,Gaensler}. These errors can, for instance, happen when measuring the magnitudes of objects in the neighborhood of very bright stars.

We have considered the mean proper motions (the absolute values) in our samples, from the square root of the sum of squares of $\mu_{ra}$ and $\mu_{dec}$ as listed in USNO. Figure \ref{Proppis} shows the proper motion distributions. We see that the mean $\mu_{total}$ is 76.7 $\pm$ 0.55 mas yr$^{-1}$ (mismatch sample) and 33.0 $\pm$ 0.51 mas yr$^{-1}$ (USNO sample). The $\mu_{total}$ differs significantly and by a factor of $\sim 2$, where the mismatch objects have higher proper motions than the typical USNO objects.

%%Finally, we considered the star/galaxy separator in the USNO, which is a classification system with integers. Here, the values range from 0 to 15. The median classification in blue is 3 (mismatch) and 3 (USNO). In red, the corresponding value is 3 (mismatch) and 4 (USNO). This indicates that the mismatch objects are equally difficult to classify as a star or galaxy, compared to a typical USNO object. 

\begin{figure*}
 \centering
   \includegraphics[scale=.5]{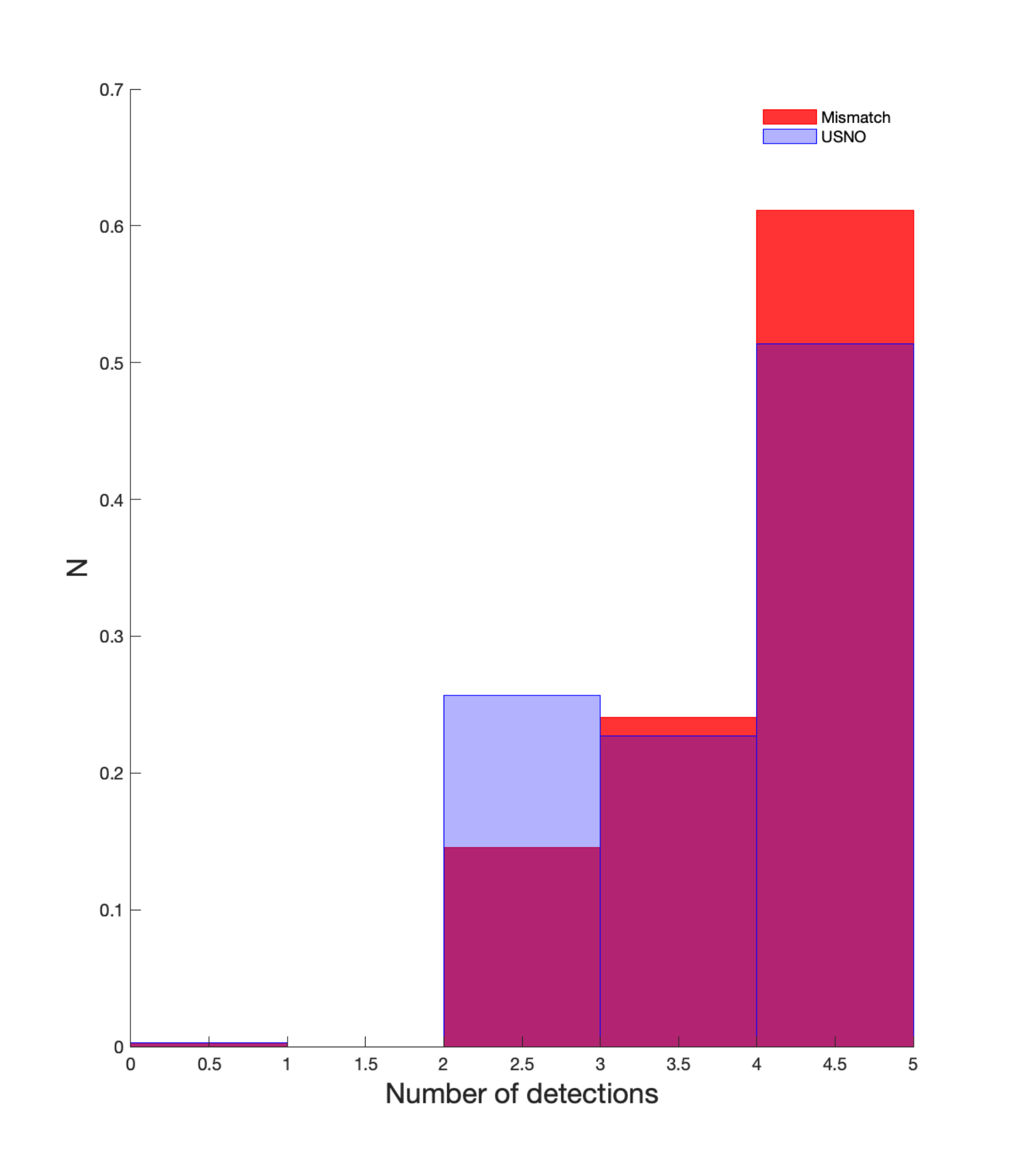}
     \caption{The total number of detections. The Mismatch Sample is compared to 
     50000 randomly-selected USNO objects. Objects in the Mismatch Sample have, on average, a higher number of detections than typical USNO objects. The histograms are normalised and zoomed.}
     \label{ndet}
     \end{figure*}

\begin{figure*}
 \centering
   \includegraphics[scale=.5]{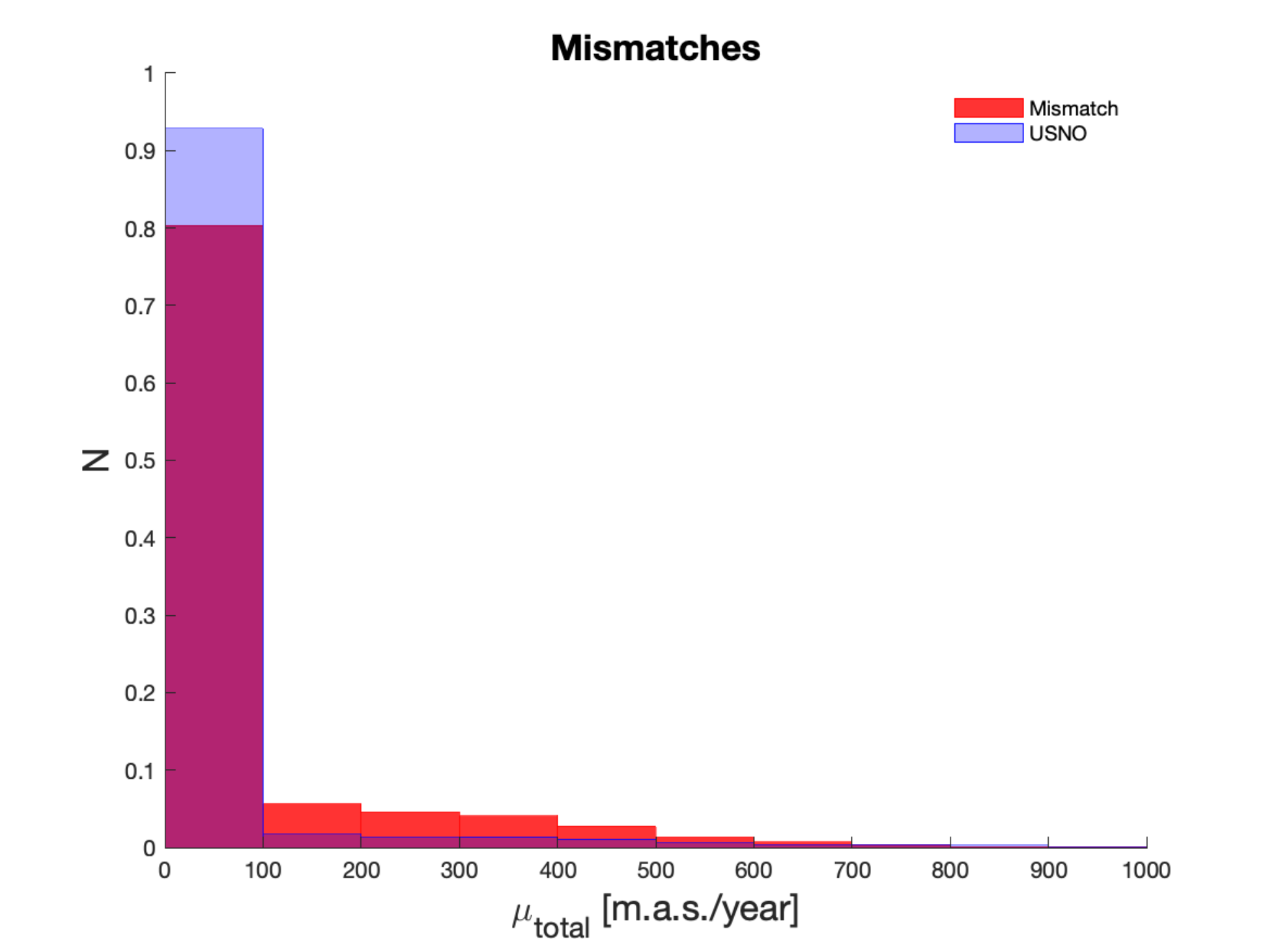}
     \caption{The USNO proper motions  $\mu_{total}$. The Mismatch Sample is compared to 50000 randomly-selected USNO objects. The histograms are normalised.}
     \label{Proppis}
     \end{figure*}

Interestingly, the objects in the Mismatch Sample show a larger number of detections, $\sim 3.8$ per object compared to the average object in the USNO sample with $\sim 3.5$ detections per object. See Fig \ref{ndet}.

Summing up, we have learned that the objects we find as mismatches are in general redder and have higher proper motions. This means nearby ($< 100$ pc) red stars could be significant contributors to the $\sim$ 150,000 Mismatch Sample. For instance, M dwarfs with magnetic flares could be among these. But what we see may also mean that the different detections for ``one'' USNO object may correspond to different objects, that happen to be close to each other by chance.

\subsection{The visually inspected sample}\label{sec:inspection}

We examine the final 1691 candidates and compare the old and new images between the DSS1 and the SDSS, complementing the study with images in several bands in the STSCI archive when the DSS1 images were not clear enough. At this stage, most of the candidates are the result
of slightly offset coordinates, and the images reveal that the objects are present in both
old and new images, with tiny offsets of the central point. About $\sim$ 200 of the 1691 candidates
are caused by dead stripes in the SDSS. Finally, about $\sim$ 100 candidates remain, most with a point-like appearance. Nearly all these candidates are single-epoch observations in the POSS1, red band. This is likely due to the way the USNO catalogue was constructed,
but possibly also due to the order of visual inspection, where we started by first examining the POSS-1 red images, later the POSS-1 blue, etc, which could introduce a bias in favour of detecting these one-time events in the POSS-1 red band.

One of possible ways of weeding out plate flaws using only the digital scans is by examining the PSFs of stars with a similar magnitude range on each plate, and comparing them to the PSF measured for each candidate. Using DS9 we measured the radial profile of the light distribution near a typical star, where the width of the PSF is estimated to be the full width at half maximum (FWHM) of the Gaussian. The sharpness of the stars varies somewhat between the plates, but when we compare the FWHM of the given transient on a plate with the FWHM of a well-known star in the same plate, we find that the full widths at half maximum are similar in most cases. However, about $\sim$ 20 objects need to be removed as they either appear asymmetric or their widths are significantly smaller than that of real stars. We find that some typical artifacts have a FWHM significantly larger than normal stars, and we therefore remove candidates with significantly larger PSFs as well. We, however, keep candidates that look like binary stars or multiple star systems, even if these could well be artifacts.

We list the $\sim$ 100 surviving objects (preliminary candidates) in Table \ref{GiantTable}. No candidate has a cross-match within 30 arc sec of a sourcein in the General Catalogue of Variable Stars \citep{Sternberg}, which means none of them are already known variable objects.

\begin{figure*}
   \centering
  \includegraphics[scale=0.1]{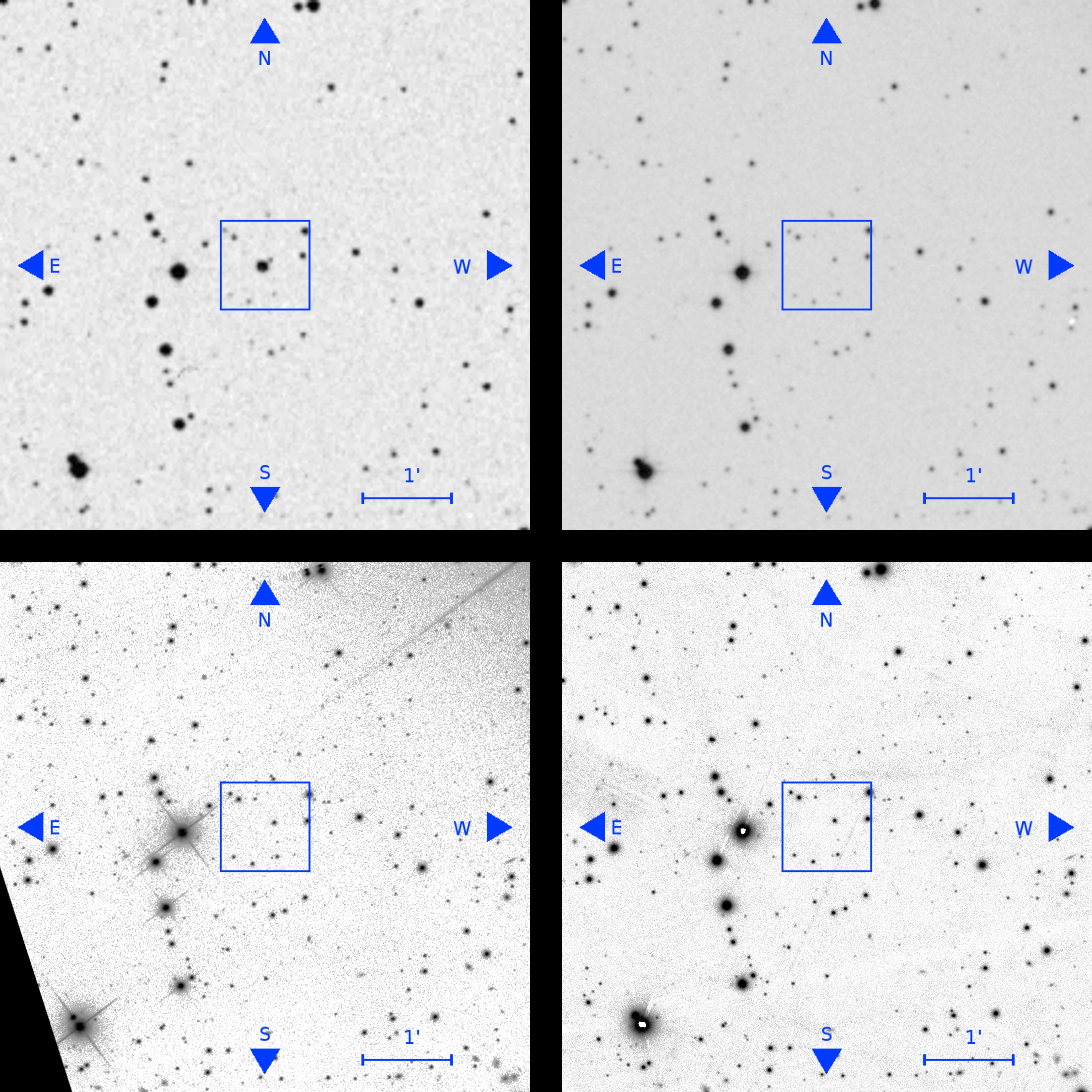}
  \caption{Example of one candidate shown in Table \ref{GiantTable}. We show the images from (upper left) POSS-1 E red, (upper right) POSS-2 red, (lower left) combination of SDSS filters, (lower right) Pan-STARRS r. The object is seen in POSS-1 red band and has the coordinates (ra,dec) = 277.7422, 40.90004. Afterwards, it seems to have``vanished''. Fine centering the coordinates of this object gives (ra,dec)=277.734042, 40.9054433.}\label{cand1}
   \end{figure*}

\begin{figure*}
  \centering
  \includegraphics[scale=0.1]{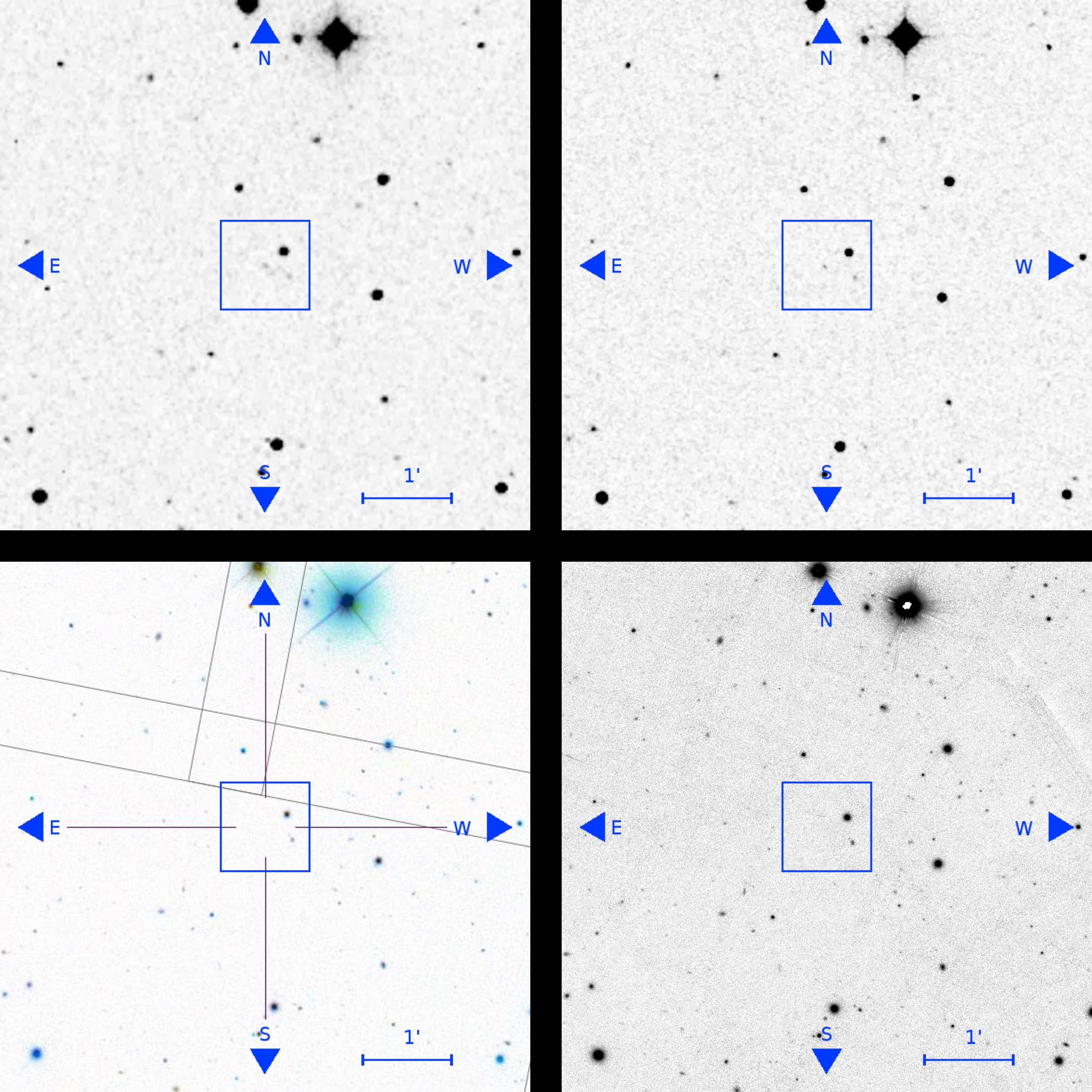}
  \caption{Note the bright star in the center in all images. We show the images from (upper left) POSS-1 E red, (upper right) POSS-2 red, (lower left) SDSS, (lower right) Pan-STARRS r. In the upper two images (POSS-1 and POSS-2 red filters), both the faint stars are present. In the lower two images (SDSS and Pan-STARRS r), only the right faint star is present.}
   \label{cand2}
   \end{figure*}

In a separate article we examine each surviving object in depth with the aim of identifying what is the true nature of our sources and to select the top candidates. As individual examples, we include images of typical objects that only are seen in one epoch. See Figures \ref{cand1} and \ref{cand2}.
The latter candidate stands out among the others. We see how something is visible in the POSS-1 and POSS-2 red filters, but with a slight shift. In the more recent images from SDSS and Pan-STARRS nothing is visible, as can be seen in the figure. However, one must take into account the exact location, the signal-to-noise ratio of the detections, the elongated fibre-like structure next to the two stars in POSS-1 (possibly an artifact?). This needs further investigation. While the blue POSS filters are not shown here, there is possibly also an extremely faint detection in the POSS-1 blue filter, but nothing at all visible in the POSS-2 blue.

   %% IMPORTANT!!! PUT THIS PICTURE BACK!!!
   
\section{Discussion}

The VASCO project aims to look for vanishing and appearing objects using old and new sky surveys. In 2016 we performed a pilot study \citep{Villarroel2016} and searched for vanishing stars in a cross-match of 10 million no-proper motion objects in USNO and SDSS (since 2000). We found one point source and established that the probability discovering vanishing events was about one in 10 million (or less) within this time frame of roughly a decade. 

We have now performed a follow-up analysis of the old images, more archival searches, and performed new observations of this object from Teide Observatory and the Nordic Optical Telescope. Near the original location, within 2.4 arcsec and 1.4 arcsec, respectively, two fainter objects that are approximately 4 to 4.5 mag fainter can be found in the red band.

We conclude that there are four possibilities: 
\begin{enumerate}
\item The detection is a variable object that has dropped approximately 4 to 4.5 mag between the 1950s and 2018.
\item The detection is a very red (or redshifted) transient event that happened in March 1950. It could have been a M dwarf that flared during the POSS-1 exposure.
\item The object is a plate scratch. This appears unlikely due to the point-like nature of the detection itself.
\item The object is a nearby, red, faint low-mass star or brown dwarf with a very high proper motion that has allowed the object to move 4.5 arcmin over a time span of 70 years. As there are few stars with such high proper motions in USNO, this appears not too likely either.
\end{enumerate}

Since only one secure detection of these objects exists, and although that detection seems is of a point source, it is difficult to establish its nature.

We have performed a new, deeper cross-match of 600 million of objects from USNO and the entire Pan-STARRS DR1 (starting in 2013) to search for more convincing vanished candidates, which supersedes the previous USNO sample by a factor of 60. As Pan-STARRS goes deeper than SDSS, the new cross-match therefore allows us to exclude a large number of variable objects near the detection limit. 
%We get a final sample of about 150 000 mismatches ('Mismatch sample') that lack a counterpart in the Pan-STARRS, that will be investigated through a citizen science project. The citizen science project may be supported by a machine-learning driven tool, see Section \ref{sec:ongoing}. In this article, however, we have investigated the properties of the Mismatch Sample, and found that the mismatches are generally redder, more variable in the red band, and have higher average proper motion.
We obtain a final sample of about 150,000 mismatches (the ``Mismatch Sample'') characterized by the lack of a counterpart in Pan-STARRS.  We have investigated the properties of the Mismatch Sample and found that the mismatches are generally redder, more variable in the red band, and have higher average proper motion. Many of these could be M dwarfs closer than  100 pc, and if an M dwarf was flaring during the POSS-1 exposure, it could be invisible in the Pan-STARRS and SDSS surveys.

Cross-matching the Mismatch Sample with the SDSS, we find 23,667 objects cannot be found in SDSS. The number of USNO objects surveyed by this cross-match is about 91 million. We examined each of these candidates in this subset visually. Most are artifacts of various sorts. However, about 100 candidates are point sources visible only in the photographic POSS-1 plates that were taken from the 1950s to 1970s. That means that the complete mismatch sample should contain at least 700 detections of this class. This is a significantly larger number than the eight known objects in our Galaxy that have proper motion larger than 5 arcsec yr$^{-1}$. A mismatch sample utilizing a 5 arcsec cross-match radius instead of the 30 arcsec cross-match radius currently used is expected to provide even more potential detections.

With the visual inspection performed on a subset of the Mismatch Sample (see Section \ref{sec:visual}) we have considered the most interesting candidates from a sample of roughly 90 million USNO objects. Among these, no truly vanishing star was convincingly detected, which means we can expect the chance of finding vanishing-star events during 70 years is less than 1 in 90 million in our Galaxy. In the Appendix we demonstrate with theoretical calculations that it is not likely to encounter a failed supernova in the VASCO searches.

\subsection{One-hundred red events?}

 %\begin{figure*}
  %% \centering
  %\includegraphics[scale=0.5]{final_colors.eps}
  %\caption{The $b - v$ lower limit of the 123 candidates.}\label{colorrrr}
  % \end{figure*}
   
What do we actually know about the transients? For the \cite{Villarroel2016} object, we can use the USNO limiting magnitude $b \sim 21$ to set a lower limit of the color, $b - r \geq$ 1.3. But many other events in the Table \ref{GiantTable} appear to have much redder colors up to $b - r \geq$ 7.4, which may mean that our objects are a mixture of \textit{apparently} red events. For some objects the red and blue observations might have happened simultaneously, while for others, there may have been a significant offset in time between the red and blue observations. We note the similarity of the \cite{Villarroel2016} object and the nuclear transient reported in Figure 6 of \cite{Djorgovski2001}, where an event with $r \sim 18.5$ is observed in its bright phase in one red image, and only seems to be ``extremely faint'' in two other filters while revealing a background galaxy at $z \sim 1$ with $r \sim 24.5$. In our case, the NOT image reveals two background objects within 1.4 and 2.4 arcsec angular distance\footnote{Note that the POSS-1 E plates resolution is 1.7 arcsec per pixel.}, close to the original spot. They too could be galaxies at high redshift. However, the offset in position
between the USNO and NOT detections that either is caused by the low resolution of POSS-1 or by proper motion puts this explanation into doubt.

Most of the 100 events could be detected in one image and not be detected again. If one assumes that most of these events were detected in two filters at the same time, they are unusually red to be Solar System objects, with half of the objects having colors $b - r \geq$ 2. Solar System objects are typically much bluer (due to the color of reflected sunlight), even if rare exceptions exist. Taking Fig. \ref{cand1} as a typical example, the POSS-1 red band and the blue band image were obtained with about a quarter hour time difference according to the listed epochs for each image in the DSS plate finder\footnote{https://archive.stsci.edu/cgi-bin/dss\_plate\_finder}. The exposure time for the red image is about 50 minutes. If the object were an asteroid and was quickly moving through the field of the red image in a few minutes of time, the object would be elongated on the plate. This object is, however, point-like. In addition, the candidate is anomalously red and not seen in the blue band, which further decreases the likelihood that it is an asteroid.

For nearby high proper motion stars to comprise most of our sample, we have far too many candidates. From the Gaia survey we know that there exist only eight stars with proper motions larger than 5 arcsec per year, which is the minimumn proper motion needed to explain the ``vanishing'' events. Therefore, it is unlikely that many of the 100 events are high proper motion objects.

Other events we may have observed are novae, supernovae at high redshift, and microlensing events or flares from M dwarfs. Some of the red transients might be intermediate-luminosity red transients (ILRTs) \citep{Bond2009} or tidal disruption events. Some of our candidates might be M dwarf flares,
as many of our objects are faint ($r \sim 18 - 19$), red and appear to have non-zero proper motion. M dwarfs tend to brighten several magnitudes during a flare, and recently a flare of 10 magnitudes was reported \citep{Rodriguez2018}.

We examined the digitised photographic plates for typical plate flaws in Section \ref{sec:inspection} and the objects listed in Table \ref{GiantTable} have satisfied the selection criteria. We propose that these objects may be worth following up with transient sky surveys to see if they may be recovered. We will analyze each of these 100 sources in a separate paper and attempt to carry out deep imaging of them. 

Following the VASCO criteria introduced at the end of the Section \ref{sec:intro}, we define the most interesting candidates as either the single-time transients with large amplitudes $\Delta m > $ 5 magnitudes, and also objects that were detected in more than one image prior to ``disappearance''. These objects are listed in Table \ref{ExtraBeautiful}. The candidates displayed in Figures \ref{cand1} and \ref{cand2} belong to this table.

\begin{table*}
	%%\centering
	\caption{\label{GiantTable}Coordinates (J2000.0) of the first set of $\sim$ 100 surviving candidates. The candidates are listed in degrees and in order of right ascension. The candidates, that have been visually identified, are not always located in the absolute centre of an image from POSS due to inaccuracies in astrometry, but often in the central region of the image. The listed red magnitudes (epoch-1) from USNO may suffer from large uncertainties due to the issues in photometry.}
%%	{\tiny
%%\setlength\tabcolsep{2pt}
\begin{center}
\begin{tabular}{c c c | c c c | c c c}
RA & Dec &$r$ (mag)&RA & Dec &$r$ (mag)&RA & Dec &$r$ (mag)\\
\hline\hline
$00^{\rm h}11^{\rm m}19.\!^{\rm s}43$&$-03^{\rm o}09^{\prime}45.\!^{\prime\prime}22$&19.18&$10^{\rm h}03^{\rm m}33.\!^{\rm s}36$&$+22^{\rm o}09^{\prime}01.\!^{\prime\prime}65$&19.27&$12^{\rm h}38^{\rm m}21.\!^{\rm s}48$&$+42^{\rm o}45^{\prime}09.\!^{\prime\prime}17$&19.26\\
$00^{\rm h}13^{\rm m}53.\!^{\rm s}05$&$+03^{\rm o}23^{\prime}20.\!^{\prime\prime}51$&19.32&$10^{\rm h}03^{\rm m}39.\!^{\rm s}53$&$+16^{\rm o}45^{\prime}01.\!^{\prime\prime}19$&17.94&$12^{\rm h}42^{\rm m}11.\!^{\rm s}06$&$+16^{\rm o}08^{\prime}55.\!^{\prime\prime}03$&13.57\\
$00^{\rm h}15^{\rm m}34.\!^{\rm s}89$&$+17^{\rm o}35^{\prime}24.\!^{\prime\prime}87$&19.09&$10^{\rm h}09^{\rm m}55.\!^{\rm s}66$&$+14^{\rm o}45^{\prime}07.\!^{\prime\prime}56$&14.45&$12^{\rm h}53^{\rm m}46.\!^{\rm s}42$&$+27^{\rm o}21^{\prime}08.\!^{\prime\prime}75$&18.13\\
$00^{\rm h}16^{\rm m}07.\!^{\rm s}23$&$+21^{\rm o}58^{\prime}47.\!^{\prime\prime}42$&18.15&$10^{\rm h}15^{\rm m}27.\!^{\rm s}12$&$+25^{\rm o}39^{\prime}43.\!^{\prime\prime}92$&19.42&$12^{\rm h}53^{\rm m}59.\!^{\rm s}30$&$+62^{\rm o}17^{\prime}01.\!^{\prime\prime}21$&19.22\\
$00^{\rm h}37^{\rm m}25.\!^{\rm s}41$&$+27^{\rm o}12^{\prime}34.\!^{\prime\prime}53$&19.36&$10^{\rm h}18^{\rm m}59.\!^{\rm s}33$&$+21^{\rm o}14^{\prime}13.\!^{\prime\prime}16$&19.38&$13^{\rm h}08^{\rm m}16.\!^{\rm s}49$&$+18^{\rm o}46^{\prime}48.\!^{\prime\prime}79$&19.10\\
$00^{\rm h}41^{\rm m}22.\!^{\rm s}82$&$+03^{\rm o}22^{\prime}04.\!^{\prime\prime}51$&19.42&$10^{\rm h}19^{\rm m}27.\!^{\rm s}62$&$+15^{\rm o}03^{\prime}17.\!^{\prime\prime}82$&18.89&$13^{\rm h}30^{\rm m}37.\!^{\rm s}34$&$+15^{\rm o}04^{\prime}19.\!^{\prime\prime}70$&19.32\\
$00^{\rm h}42^{\rm m}38.\!^{\rm s}75$&$+12^{\rm o}30^{\prime}47.\!^{\prime\prime}66$&18.40&$10^{\rm h}22^{\rm m}04.\!^{\rm s}56$&$+24^{\rm o}26^{\prime}28.\!^{\prime\prime}03$&18.90&$13^{\rm h}30^{\rm m}54.\!^{\rm s}17$&$+02^{\rm o}30^{\prime}07.\!^{\prime\prime}78$&19.05\\
$00^{\rm h}51^{\rm m}15.\!^{\rm s}20$&$-05^{\rm o}40^{\prime}23.\!^{\prime\prime}84$&19.42&$10^{\rm h}23^{\rm m}32.\!^{\rm s}71$&$+16^{\rm o}50^{\prime}09.\!^{\prime\prime}06$&18.88&$13^{\rm h}32^{\rm m}01.\!^{\rm s}75$&$+12^{\rm o}10^{\prime}19.\!^{\prime\prime}31$&14.64\\
$00^{\rm h}58^{\rm m}32.\!^{\rm s}18$&$+17^{\rm o}48^{\prime}04.\!^{\prime\prime}25$&19.41&$10^{\rm h}30^{\rm m}27.\!^{\rm s}43$&$+22^{\rm o}44^{\prime}17.\!^{\prime\prime}16$&19.08&$13^{\rm h}45^{\rm m}29.\!^{\rm s}57$&$+27^{\rm o}44^{\prime}12.\!^{\prime\prime}26$&19.36\\
$01^{\rm h}39^{\rm m}29.\!^{\rm s}57$&$+09^{\rm o}00^{\prime}39.\!^{\prime\prime}28$&19.38&$10^{\rm h}40^{\rm m}48.\!^{\rm s}43$&$+21^{\rm o}53^{\prime}28.\!^{\prime\prime}36$&19.36&$13^{\rm h}55^{\rm m}01.\!^{\rm s}30$&$+08^{\rm o}10^{\prime}42.\!^{\prime\prime}56$&19.36\\
$01^{\rm h}55^{\rm m}30.\!^{\rm s}70$&$+09^{\rm o}00^{\prime}42.\!^{\prime\prime}19$&18.82&$10^{\rm h}43^{\rm m}14.\!^{\rm s}40$&$+17^{\rm o}10^{\prime}52.\!^{\prime\prime}50$&18.99&$13^{\rm h}55^{\rm m}06.\!^{\rm s}31$&$+11^{\rm o}11^{\prime}46.\!^{\prime\prime}36$&17.51\\
$01^{\rm h}56^{\rm m}36.\!^{\rm s}11$&$+22^{\rm o}52^{\prime}08.\!^{\prime\prime}83$&16.91&$10^{\rm h}51^{\rm m}35.\!^{\rm s}73$&$+15^{\rm o}13^{\prime}38.\!^{\prime\prime}71$&18.91&$14^{\rm h}10^{\rm m}38.\!^{\rm s}04$&$+26^{\rm o}00^{\prime}33.\!^{\prime\prime}45$&19.02\\
$02^{\rm h}12^{\rm m}54.\!^{\rm s}67$&$+06^{\rm o}51^{\prime}45.\!^{\prime\prime}29$&16.24&$10^{\rm h}55^{\rm m}38.\!^{\rm s}33$&$+14^{\rm o}55^{\prime}38.\!^{\prime\prime}75$&18.54&$14^{\rm h}11^{\rm m}43.\!^{\rm s}63$&$+26^{\rm o}49^{\prime}39.\!^{\prime\prime}07$&18.10\\
$02^{\rm h}53^{\rm m}01.\!^{\rm s}89$&$-01^{\rm o}40^{\prime}11.\!^{\prime\prime}68$&18.54&$11^{\rm h}07^{\rm m}51.\!^{\rm s}02$&$+35^{\rm o}02^{\prime}14.\!^{\prime\prime}50$&19.23&$14^{\rm h}11^{\rm m}57.\!^{\rm s}26$&$+12^{\rm o}33^{\prime}49.\!^{\prime\prime}79$&18.66\\
$03^{\rm h}31^{\rm m}34.\!^{\rm s}89$&$+05^{\rm o}24^{\prime}46.\!^{\prime\prime}04$&18.47&$11^{\rm h}07^{\rm m}53.\!^{\rm s}57$&$+18^{\rm o}50^{\prime}28.\!^{\prime\prime}18$&18.38&$14^{\rm h}22^{\rm m}50.\!^{\rm s}45$&$+44^{\rm o}21^{\prime}32.\!^{\prime\prime}54$&18.90\\
$04^{\rm h}31^{\rm m}50.\!^{\rm s}68$&$+08^{\rm o}16^{\prime}14.\!^{\prime\prime}63$&19.23&$11^{\rm h}14^{\rm m}45.\!^{\rm s}24$&$+08^{\rm o}17^{\prime}46.\!^{\prime\prime}46$&18.23&$14^{\rm h}25^{\rm m}23.\!^{\rm s}11$&$+32^{\rm o}53^{\prime}55.\!^{\prime\prime}35$&19.23\\
$04^{\rm h}43^{\rm m}56.\!^{\rm s}87$&$+11^{\rm o}45^{\prime}37.\!^{\prime\prime}48$&18.54&$11^{\rm h}20^{\rm m}13.\!^{\rm s}49$&$+07^{\rm o}56^{\prime}55.\!^{\prime\prime}72$&18.69&$14^{\rm h}46^{\rm m}00.\!^{\rm s}91$&$+32^{\rm o}12^{\prime}45.\!^{\prime\prime}43$&19.39\\
$06^{\rm h}38^{\rm m}49.\!^{\rm s}26$&$+82^{\rm o}51^{\prime}04.\!^{\prime\prime}42$&15.12&$11^{\rm h}24^{\rm m}26.\!^{\rm s}11$&$+02^{\rm o}07^{\prime}13.\!^{\prime\prime}76$&19.03&$14^{\rm h}49^{\rm m}47.\!^{\rm s}57$&$+62^{\rm o}34^{\prime}42.\!^{\prime\prime}46$&19.47\\
$07^{\rm h}47^{\rm m}37.\!^{\rm s}80$&$+37^{\rm o}21^{\prime}22.\!^{\prime\prime}64$&18.57&$11^{\rm h}31^{\rm m}17.\!^{\rm s}98$&$+10^{\rm o}13^{\prime}25.\!^{\prime\prime}36$&18.60&$15^{\rm h}04^{\rm m}27.\!^{\rm s}93$&$+62^{\rm o}23^{\prime}24.\!^{\prime\prime}22$&18.11\\
$08^{\rm h}01^{\rm m}50.\!^{\rm s}42$&$+24^{\rm o}36^{\prime}03.\!^{\prime\prime}24$&19.01&$11^{\rm h}31^{\rm m}34.\!^{\rm s}90$&$+04^{\rm o}15^{\prime}51.\!^{\prime\prime}55$&19.25&$15^{\rm h}12^{\rm m}07.\!^{\rm s}44$&$+51^{\rm o}06^{\prime}36.\!^{\prime\prime}07$&18.83\\
$08^{\rm h}21^{\rm m}44.\!^{\rm s}71$&$+58^{\rm o}18^{\prime}12.\!^{\prime\prime}27$&19.01&$11^{\rm h}32^{\rm m}16.\!^{\rm s}08$&$+02^{\rm o}34^{\prime}24.\!^{\prime\prime}31$&19.03&$15^{\rm h}22^{\rm m}30.\!^{\rm s}96$&$+24^{\rm o}40^{\prime}05.\!^{\prime\prime}99$&17.62\\
$08^{\rm h}22^{\rm m}16.\!^{\rm s}49$&$+37^{\rm o}46^{\prime}05.\!^{\prime\prime}12$&18.27&$11^{\rm h}32^{\rm m}25.\!^{\rm s}37$&$+05^{\rm o}22^{\prime}57.\!^{\prime\prime}00$&19.21&$15^{\rm h}29^{\rm m}41.\!^{\rm s}43$&$+22^{\rm o}58^{\prime}18.\!^{\prime\prime}19$&16.18\\
$08^{\rm h}29^{\rm m}44.\!^{\rm s}74$&$+61^{\rm o}29^{\prime}34.\!^{\prime\prime}84$&18.95&$11^{\rm h}49^{\rm m}31.\!^{\rm s}70$&$+16^{\rm o}07^{\prime}17.\!^{\prime\prime}33$&17.05&$15^{\rm h}38^{\rm m}31.\!^{\rm s}44$&$+43^{\rm o}02^{\prime}04.\!^{\prime\prime}96$&17.57\\
$08^{\rm h}35^{\rm m}18.\!^{\rm s}72$&$+62^{\rm o}03^{\prime}51.\!^{\prime\prime}51$&19.43&$11^{\rm h}53^{\rm m}31.\!^{\rm s}90$&$+63^{\rm o}03^{\prime}59.\!^{\prime\prime}36$&19.43&$15^{\rm h}41^{\rm m}04.\!^{\rm s}78$&$+09^{\rm o}00^{\prime}54.\!^{\prime\prime}50$&16.49\\
$08^{\rm h}44^{\rm m}04.\!^{\rm s}68$&$+57^{\rm o}56^{\prime}55.\!^{\prime\prime}54$&17.84&$11^{\rm h}56^{\rm m}22.\!^{\rm s}08$&$+67^{\rm o}04^{\prime}36.\!^{\prime\prime}33$&19.49&$15^{\rm h}48^{\rm m}55.\!^{\rm s}08$&$+62^{\rm o}24^{\prime}52.\!^{\prime\prime}38$&19.25\\
$08^{\rm h}51^{\rm m}17.\!^{\rm s}30$&$+62^{\rm o}58^{\prime}36.\!^{\prime\prime}16$&18.91&$12^{\rm h}17^{\rm m}49.\!^{\rm s}03$&$+67^{\rm o}40^{\prime}27.\!^{\prime\prime}40$&19.27&$16^{\rm h}09^{\rm m}45.\!^{\rm s}19$&$+38^{\rm o}34^{\prime}48.\!^{\prime\prime}68$&18.85\\
$09^{\rm h}18^{\rm m}46.\!^{\rm s}22$&$+62^{\rm o}29^{\prime}49.\!^{\prime\prime}10$&18.60&$12^{\rm h}23^{\rm m}15.\!^{\rm s}99$&$+16^{\rm o}28^{\prime}14.\!^{\prime\prime}92$&19.23&$16^{\rm h}10^{\rm m}01.\!^{\rm s}20$&$+61^{\rm o}31^{\prime}11.\!^{\prime\prime}25$&17.79\\
$09^{\rm h}33^{\rm m}09.\!^{\rm s}84$&$+15^{\rm o}08^{\prime}35.\!^{\prime\prime}99$&19.17&$12^{\rm h}29^{\rm m}37.\!^{\rm s}58$&$+61^{\rm o}17^{\prime}59.\!^{\prime\prime}97$&18.44&$16^{\rm h}24^{\rm m}00.\!^{\rm s}31$&$+61^{\rm o}21^{\prime}00.\!^{\prime\prime}26$&18.21\\
$09^{\rm h}36^{\rm m}21.\!^{\rm s}19$&$+37^{\rm o}49^{\prime}43.\!^{\prime\prime}86$&18.82&$12^{\rm h}30^{\rm m}28.\!^{\rm s}80$&$+20^{\rm o}43^{\prime}00.\!^{\prime\prime}95$&16.95&$16^{\rm h}34^{\rm m}25.\!^{\rm s}99$&$+34^{\rm o}04^{\prime}24.\!^{\prime\prime}45$&17.40\\
$09^{\rm h}39^{\rm m}28.\!^{\rm s}66$&$+22^{\rm o}39^{\prime}41.\!^{\prime\prime}73$&19.48&$12^{\rm h}31^{\rm m}55.\!^{\rm s}13$&$-02^{\rm o}06^{\prime}22.\!^{\prime\prime}03$&18.24&$17^{\rm h}20^{\rm m}34.\!^{\rm s}85$&$+44^{\rm o}00^{\prime}11.\!^{\prime\prime}34$&19.49\\
$09^{\rm h}42^{\rm m}00.\!^{\rm s}89$&$+16^{\rm o}55^{\prime}56.\!^{\prime\prime}32$&18.86&$12^{\rm h}33^{\rm m}09.\!^{\rm s}27$&$+43^{\rm o}14^{\prime}07.\!^{\prime\prime}05$&19.37&$17^{\rm h}25^{\rm m}44.\!^{\rm s}62$&$+73^{\rm o}54^{\prime}09.\!^{\prime\prime}76$&19.39\\
$09^{\rm h}43^{\rm m}07.\!^{\rm s}39$&$+31^{\rm o}11^{\prime}59.\!^{\prime\prime}38$&18.77&$12^{\rm h}37^{\rm m}23.\!^{\rm s}88$&$+05^{\rm o}26^{\prime}29.\!^{\prime\prime}40$&19.33&$18^{\rm h}30^{\rm m}58.\!^{\rm s}12$&$+40^{\rm o}54^{\prime}00.\!^{\prime\prime}14$&18.55\\
$09^{\rm h}50^{\rm m}13.\!^{\rm s}78$&$+49^{\rm o}39^{\prime}16.\!^{\prime\prime}09$&18.32&$12^{\rm h}37^{\rm m}53.\!^{\rm s}42$&$+12^{\rm o}55^{\prime}34.\!^{\prime\prime}50$&17.29&$18^{\rm h}53^{\rm m}29.\!^{\rm s}81$&$+77^{\rm o}55^{\prime}00.\!^{\prime\prime}54$&17.23\\
\hline
\end{tabular}
\end{center}
\end{table*}

\begin{table}
	%%\centering
	\caption{\label{ExtraBeautiful}The most interesting candidates. For the events listed in Table \ref{GiantTable}, we remeasured the coordinates of the interesting candidates. The list contains all events showing a single point source with $r$ < 18.4, either as measured by the listed USNO magnitudes or when we remeasured its magnitude directly from the digitalized plates. Also one object that appears to be seen in more than one image, is included.}
%%	{\tiny
%%\setlength\tabcolsep{2pt}
\begin{center}
\begin{tabular}{c c| c c}
RA & Dec &RA & Dec \\
\hline\hline
$00^{\rm h}16^{\rm m}06.\!^{\rm s}92$&$+21^{\rm o}58^{\prime}47.\!^{\prime\prime}82$&$11^{\rm h}56^{\rm m}18.\!^{\rm s}10$&$+67^{\rm o}04^{\prime}23.\!^{\prime\prime}15$\\
$00^{\rm h}42^{\rm m}38.\!^{\rm s}77$&$+12^{\rm o}30^{\prime}36.\!^{\prime\prime}75$&$12^{\rm h}31^{\rm m}55.\!^{\rm s}05$&$-02^{\rm o}06^{\prime}28.\!^{\prime\prime}11$\\
$00^{\rm h}58^{\rm m}32.\!^{\rm s}63$&$+17^{\rm o}48^{\prime}19.\!^{\prime\prime}44$&$12^{\rm h}42^{\rm m}11.\!^{\rm s}02$&$+16^{\rm o}08^{\prime}57.\!^{\prime\prime}25$\\
$01^{\rm h}56^{\rm m}36.\!^{\rm s}27$&$+22^{\rm o}52^{\prime}11.\!^{\prime\prime}87$&$12^{\rm h}53^{\rm m}46.\!^{\rm s}58$&$+27^{\rm o}21^{\prime}06.\!^{\prime\prime}67$\\
$02^{\rm h}12^{\rm m}54.\!^{\rm s}59$&$+06^{\rm o}51^{\prime}42.\!^{\prime\prime}78$&$12^{\rm h}53^{\rm m}55.\!^{\rm s}05$&$+62^{\rm o}17^{\prime}08.\!^{\prime\prime}94$\\
$02^{\rm h}53^{\rm m}02.\!^{\rm s}20$&$-01^{\rm o}40^{\prime}19.\!^{\prime\prime}44$&$13^{\rm h}32^{\rm m}01.\!^{\rm s}90$&$+12^{\rm o}10^{\prime}17.\!^{\prime\prime}53$\\
$08^{\rm h}22^{\rm m}17.\!^{\rm s}55$&$+37^{\rm o}46^{\prime}20.\!^{\prime\prime}67$&$15^{\rm h}38^{\rm m}31.\!^{\rm s}31$&$+43^{\rm o}02^{\prime}01.\!^{\prime\prime}23$\\
$09^{\rm h}39^{\rm m}28.\!^{\rm s}81$&$+22^{\rm o}39^{\prime}40.\!^{\prime\prime}76$&$15^{\rm h}41^{\rm m}05.\!^{\rm s}11$&$+09^{\rm o}00^{\prime}54.\!^{\prime\prime}75$\\
$09^{\rm h}50^{\rm m}13.\!^{\rm s}97$&$+49^{\rm o}39^{\prime}36.\!^{\prime\prime}40$&$15^{\rm h}48^{\rm m}54.\!^{\rm s}99$&$+62^{\rm o}24^{\prime}48.\!^{\prime\prime}04$\\
$10^{\rm h}03^{\rm m}39.\!^{\rm s}48$&$+16^{\rm o}44^{\prime}55.\!^{\prime\prime}61$&$16^{\rm h}10^{\rm m}01.\!^{\rm s}70$&$+61^{\rm o}31^{\prime}06.\!^{\prime\prime}88$\\
$10^{\rm h}55^{\rm m}38.\!^{\rm s}50$&$+14^{\rm o}55^{\prime}29.\!^{\prime\prime}15$&$16^{\rm h}24^{\rm m}00.\!^{\rm s}44$&$+61^{\rm o}20^{\prime}59.\!^{\prime\prime}43$\\
$11^{\rm h}07^{\rm m}51.\!^{\rm s}05$&$+35^{\rm o}02^{\prime}12.\!^{\prime\prime}17$&$16^{\rm h}34^{\rm m}26.\!^{\rm s}23$&$+34^{\rm o}04^{\prime}25.\!^{\prime\prime}11$\\
$11^{\rm h}24^{\rm m}24.\!^{\rm s}06$&$+02^{\rm o}07^{\prime}26.\!^{\prime\prime}26$&$18^{\rm h}30^{\rm m}56.\!^{\rm s}35$&$+40^{\rm o}54^{\prime}17.\!^{\prime\prime}80$\\
$11^{\rm h}49^{\rm m}31.\!^{\rm s}87$&$+16^{\rm o}07^{\prime}15.\!^{\prime\prime}09$&$18^{\rm h}53^{\rm m}29.\!^{\rm s}27$&$+77^{\rm o}54^{\prime}56.\!^{\prime\prime}64$\\
\hline
\end{tabular}

\end{center}
\end{table}

\subsection{Implications for SETI research}\label{sec:seti}

The Search for Extraterrestrial Intelligence (SETI) nowadays includes a broad set of activities, where the two large domains of searches are done both in the radio and in the optical, in hopes of finding so-called ``technosignatures'' like those we ourselves are already capable of producing, such as interstellar communication with lasers \citep{Townes}. Optical SETI searches looking for lasers are particularly interesting, as these signatures often have a low temporal dispersion (as opposed to the radio searches), and earthlings already have the necessary technology to produce short, nanosecond laser pulses that could outshine our own Sun by a factor of $\sim$ 5000. 
SETI programs such as the ``Panoramic optical and near-infrared SETI instrument'' (PANOSETI) are presently preparing instrumentation to search for short light pulses on timescales of nanoseconds to microseconds that may arise due to interstellar communication \citep{Maire2018,Wright2018}.

A number of ongoing optical search programs have already succeeded in producing some upper limits to the incidence of both pulsed lasers and continuous laser signals. For example, for 800 nm lasers, one estimates the fraction of transmitting civilisations to be around $f\sim 10^{-7}$ \citep{Technosignatures} for 100 second long pulses. A study that instead looked for signs of a continuous laser in the optical spectra of 5600 FGKM stars \citep{Marcy2015,Marcy2017} could also exclude the presence of lasers in all of these spectra. Similar searches have been done in the infrared, where extinction is much less of a problem in these wavelengths and a wavelength window is opened up that is largely devoid of background noise. See \cite{WrightNIR}.

The VASCO project may be a ``conventional'' astrophysics project, but it originated in the context of SETI, as described by \cite{Villarroel2016}, who proposed to search surveys for vanished stars in our Galaxy as probes of ``impossible effects'' that could only be ascribed to an extraterrestrial technology due to the high likelihood of this as an observational signature. While VASCO attempts to search for more transients of also natural astrophysical origins, the project bears implications for SETI research. A general review describing the possibilities of technosignature searches in the time-domain astronomy is given by \cite{Davenport2019}.

In the VASCO searches we may search for vanished stars and can expect to find transients on three different time scales: (1) a hypothetical vanished star may have existed for billions of years before it vanishes. We have determined that the probability is less than $p < 10^{-7}$. (2) We may find extreme, variable astrophysical objects that vary over timescales of decades. (3) We may find astrophysical transients that are as short as the exposure time of a typical POSS image ($\sim$ 1 hour). The short transient detections we see in the red plates from POSS-I, may have many different explanations, ranging from instrumental causes to bona fide astrophysical ones. An attractive feature about the list we have produced is that a monochromatic interstellar laser at 600 to 680 nm that shines for about one hour may well present itself as a point source only detected once in one image, due to the short time when the laser operated. Simply put, the single events presented in Section \ref{sec:visual} have many degenerate solutions.  It will be the work of a future publication to work out and disentangle this.

In SETI, frequent technosignature searches also include searches for giant structures that harness the energy of stars and produce waste heat with temperatures $T \sim 100 - 300$ K. The most extreme form is referred to as a Dyson sphere \citep{Dyson1960}, which entirely encloses a star and produces the largest fractional change in the brightness of the object. \cite{Carrigan2009} sought Dyson spheres around 11000 stars using IRAS photometry and spectroscopy. \cite{Zackrisson2015} surveyed 1359 galaxies with the help of the Tully-Fischer relation, found no convincing candidates, and estimated that the fraction of Kardashev II - III civilisations \citep{Kardashev} capable of transforming their entire galaxy is less than 0.3 percent. \cite{Griffith2015} used WISE and 2MASS to search for IR excesses amongst 100,000 targets that appear to be dust rich, star-forming galaxies. As the waste heat shows the same signatures that dust shows, for extragalactic objects these searches may be too ambiguous to give confirmable candidates. %%Moreover, the infrared excess expected from Dyson spheres, is a typical signature of the obscuring dust typical in the absolute centers of active galactic nuclei (AGN).

One may wonder why a highly advanced Kardashev II-III civilisation, capable of putting Dyson spheres around every star in a galaxy, would limit their effort to harness the energy of stars over such a giant volume as an entire galaxy. Indeed, an AGN occupies a much smaller space (as small as our Solar System), and has much more concentrated energy to offer. For example, the quasar 3C 273 has about 4 trillion times the luminosity of our Sun. Indeed, an AGN may be a significantly more effective target to build a Dyson sphere around. Many AGN (in particular obscured ones) naturally have a thick layer of dust dimming the central power source and giving off infrared emission. This dust is located at the sublimation radius. When an AGN is so obscured that hardly any photons leak through to excite the surrounding gas, we may not even detect the typical narrow emission lines that are the signature of an AGN.

When the accretion disk varies and changes its intensity, we expect the corresponding hot dust emission (arising typically $\sim$ 0.1 parsec from the supermassive black hole) to respond, but with a time delay. This time delay is often used to infer the physical size of the black hole. Together with the angular size of the torus, obtainable from interferometry, one can estimate the distance to the AGN using it as a standard candle \citep{Hoenig2014}.

However, in a dynamic, Dysonian AGN one may expect that the time delay of the infrared emission does not follow the typical behavior of a dust torus. It could be that the AGN cannot even be used as a standard candle, as the artificial structure will not obey natural changes in the power source. Therefore, as an extension of the VASCO project, we suggest searches for extra-galactic objects with variability in the infrared region. These variable AGN can be followed up with IR reverberation mapping experiments. The research will mainly be aimed at understanding the mysterious nature of the central few parsecs of an AGN.
\\
\\
\\
\subsection{Future work}
\subsubsection{Giving extraterrestrials a second chance}\label{sec:second}

Undoubtedly, VASCO will generate large lists of candidate objects in searches for vanishing stars. Individually, these serve no purpose unless verified. We can agree that a wide-field search that results in a list of candidates is of no great interest for research if each candidate sooner or later gets dismissed due to lack of verification as a potential SETI candidate.

However, if a region of the sky has a tendency to produce an unexpectedly large fraction of candidates relative to the background, this region or ``hot spot'' may deserve some extra attention. As a part of VASCO's research program, we plan to combine all the unverified initial results from many different search programs such as the optical all-sky surveys NIROSETI and PANOSETI, and from other wide-field surveys in general (see Section \ref{sec:seti}.  We aim to visualize the background of the unverified candidates in a two-dimensional projection of the sky. Altogether, this noisy background of neglected candidates could reveal ``hot spots'' of transient activity, where for some reason many candidates are concentrated. Doing this iteratively with reliable clustering methods and zooming in on the most active regions in our SETI (or technosignature) searches, we can identify the most probable locations to host extra-terrestrial intelligence. VASCO will therefore never dismiss any candidates forever. Rejection and acceptance are only transient states in the process. The information on potential ``hot spots'' can further be used to select the most interesting candidates.

\subsubsection{Expanding the set of candidates}\label{sec:ongoing}

What we have presented so far is a cross-match between USNO and Pan-STARRS in searches for vanishing objects, using a 30 arcsec cross-match radius. However, the plan of VASCO is to do the following:

\begin{enumerate}
\item Finalize the current search for vanishing objects with a 30 arcsec cross-match radius by examining the entire Mismatch Sample visually and finalize the cross-match over the sky regions that so far have not been used.
\item Search for appearing objects within a 30 arcsec cross-match radius (with Pan-STARRS objects having $r < 19$).
\item Search for vanishing objects within a 5 arcsec cross-match radius (setting limits in magnitude), including proper motion corrections. Given the larger number of spurious mismatches with this search radius, we will need to develop a better automatic methods to handle the identification of candidates in images.
\end{enumerate}

This is a long-winded process requiring considerable time on powerful computing clusters, but it may generate a large list of interesting transients of all sorts.

%The large number of images we are dealing with within the complete VASCO project and increased complexity of our searches, requires a better approach than was done in the pilot study. Clearly, we must explore ways to transfer as much effort as possible to automatized procedures, but without putting all the quality control responsibility onto machines. Here machine learning could help.
The large number of images we are dealing with within the complete VASCO project and the increased complexity of our searches requires a better approach than was done in the pilot study. Clearly, we must explore ways to transfer avail ourselves of automatized procedures as much as possible, but without relying on algorithms for all candidate selection and quality control. At this moment, such algorithms are still being developed. Current problems are related to the inefficiency of comparing two images manually, comparing images based on CCD cameras with images from old photographic plates, and finally we must adjust the algorithms to identify the most meaningful candidates. In a separate paper by \cite{Pelckmans} we propose a new tool for handling a large number of images using methods of machine learning.

\subsection{Summary}\label{sec:future}

VASCO is a project that provides an opportunity to discover many past transients events, both objects that vanish and those that appear. The time span between these surveys is large.  This allows for other phenomena to be discovered other than what can be expected in ongoing transient surveys like e.g. ZTF. Using a large cross-match radius of 30 arcseconds, we obtained a sample of 150,000 USNO objects that cannot to be found in Pan-STARRS. This represents an interesting starting sample in searches for vanishing objects. As we used a large cross-match radius of 30 arcsec (instead of the more typical 3 to 5 arcsec radius), we underestimate the real number of potential mismatches that can be found through cross-matching attempts. We have investigated the statistical properties of this sample and found that many of these ``mismatches'' are occurring in the red band. Visual checks confirm that indeed the most interesting cases, about 100, are mostly one time detections in the red band. At present, we do not know what these detections represent. We believe they may be a mixed bag of transient phenomena. The object found by \cite{Villarroel2016} is of the same class, and might possibly be a variable object that dropped 4.5 mag since it was imaged long ago. It could also have been some type of transient event such as a background high-redshift supernova or a flaring M dwarf. %%It will be the mission of our work in the near future to examine the 120 detections under a microscope, and to redo the photometric analysis and astrometry for the remaining, real candidates.

In good agreement with theoretical predictions for the number of failed supernovae in our Galaxy (see Appendix, \ref{sec:supernova}), we also set an upper limit on the probability in detecting a vanishing star to be less than 1 in 90 million during our time window of 70 years.

Meanwhile, we will keep developing methods to analyze the remaining images in the Mismatch Sample in searches for reliable examples of vanishing stars.

%% Finalize with Malmquist bias.

%%\section{Sample}\label{sec:Sample}
%%\begin{figure*}
%%	\includegraphics[scale=0.52]{Flowchart7may.png}
  %%  \caption{Flowchart demonstrating the overall sample selection, as described in Section \ref{sec:Sample}.}
 %%   \label{Flowchart}
%%\end{figure*}

%%\begin{table*}[t]
%%\caption{The main sample. The table shows all SN events that have been found in Seyfert-1 AGN. Names, morphologies, host galaxy redshifts $z$, SN discovery references and types are listed. Also listed is the angular distance from supernova to the galaxy centre.}
%%\centering
%%{\tiny
%%\setlength\tabcolsep{2pt}
%%\begin{tabular}{|c c c c c c c c}
%%\hline\hline
%%\multicolumn{8}{c}{Collected sample} \\
%%\hline\hline
%%Galaxy name / ID & Activity class & z & SN name &  SN discovery & SN type & SN z  & d (arcmin)\\

%%\hline
%%\hline
%%\end{tabular}
%%}
%%\label{SupernovaCollection}
%%\end{table*}

%%\newpage

\section*{Acknowledgements}

Based on observations made with the Nordic Optical Telescope operated on the island of La Palma by NOTSA in the Spanish Observatorio del Roque de los Muchachos of the Instituto de Astrofísica de Canarias under the joint IACNordic Observing Time programme. This article is also based on observations made with the IAC-80 telescope operated on the island of Tenerife by the Instituto de Astrofisica de Canarias in the Spanish Observatorio del Teide.

The cross-match computations were performed on resources provided by the Swedish National Infrastructure for Computing (SNIC) at UPPMAX. Salman Toor assisted in operating the resources.

B.V. wishes to thank Geoffrey Marcy for many great ideas, discussions and suggestions that improved the project and the manuscript. She thanks the referee for many thoughtful and constructive suggestions during the review process. %%She expresses regret for that the Nobel committee excluded G.M. from the Nobel prize in Physics 2019. 
B.V. wishes to thank Anders Nyholm for good discussions, and also thank Kjell Lundgren, Lars Hermansson and Abel Souza. A particular thanks to Nidia Morell who helped with the Magellan images. B.V. is funded by the Swedish Research Council (Vetenskapsr\aa det, grant no. 2017-06372).

We wish to thank the Department of information technology at Uppsala University for supporting this project. We further wish to thank Bernie Shiao from the Pan-STARRS collaboration for the offline Pan-STARRS DR1 dataset on which all VASCO work is built and Breakthrough Initiatives for support during the VASCO workshop 2018.

The Pan-STARRS1 Surveys (PS1) and the PS1 public science archive have been made possible through contributions by the Institute for Astronomy, the University of Hawaii, the Pan-STARRS Project Office, the Max-Planck Society and its participating institutes, the Max Planck Institute for Astronomy, Heidelberg and the Max Planck Institute for Extraterrestrial Physics, Garching, The Johns Hopkins University, Durham University, the University of Edinburgh, Queen's University Belfast, the Harvard-Smithsonian Center for Astrophysics, the Las Cumbres Observatory Global Telescope Network Incorporated, the National Central University of Taiwan, the Space Telescope Science Institute, the National Aeronautics and Space Administration under Grant No. NNX08AR22G issued through the Planetary Science Division of the NASA Science Mission Directorate, National Science Foundation Grant No. AST-1238877, the University of Maryland, Eotvos Lorand University (ELTE), the Los Alamos National Laboratory, and the Gordon and Betty Moore Foundation.

Funding for the Sloan Digital Sky Survey IV has been provided by the Alfred P. Sloan Foundation, the U.S. Department of Energy Office of Science, and the Participating Institutions. SDSS-IV acknowledges
support and resources from the Center for High-Performance Computing at
the University of Utah. The SDSS web site is www.sdss.org.

SDSS-IV is managed by the Astrophysical Research Consortium for the 
Participating Institutions of the SDSS Collaboration including the 
Brazilian Participation Group, the Carnegie Institution for Science, 
Carnegie Mellon University, the Chilean Participation Group, the French Participation Group, Harvard-Smithsonian Center for Astrophysics, Instituto de Astrof\'isica de Canarias, The Johns Hopkins University, Kavli Institute for the Physics and Mathematics of the Universe (IPMU) / University of Tokyo, the Korean Participation Group, Lawrence Berkeley National Laboratory, 
Leibniz Institut f\"ur Astrophysik Potsdam (AIP),  
Max-Planck-Institut f\"ur Astronomie (MPIA Heidelberg), 
Max-Planck-Institut f\"ur Astrophysik (MPA Garching), 
Max-Planck-Institut f\"ur Extraterrestrische Physik (MPE), 
National Astronomical Observatories of China, New Mexico State University, New York University, University of Notre Dame, 
Observat\'ario Nacional / MCTI, The Ohio State University, 
Pennsylvania State University, Shanghai Astronomical Observatory, 
United Kingdom Participation Group, Universidad Nacional Aut\'onoma de M\'exico, University of Arizona, University of Colorado Boulder, University of Oxford, University of Portsmouth, University of Utah, University of Virginia, University of Washington, University of Wisconsin, Vanderbilt University, and Yale University.

%%%%%%%%%%%%%%%%%%%%%%%%%%%%%%%%%%%%%%%%%%%%%%%%%%



%\bibliographystyle{mnras}

\begin{thebibliography}{99}

\bibitem[\protect\citeauthoryear{Abazajian et al.}{2009}]{Abazajian2009}
Abazajian, K., 2009, ApJ Supplement Series, 182, 543

\bibitem[\protect\citeauthoryear{Adams et al.}{2016}]{Adams2016}
Adams S. M., Kochanek C. S., Prieto J. L., Dai X., Shappee B. J., Stanek
K. Z., 2016, MNRAS, 460, 1645

\bibitem[\protect\citeauthoryear{Adams et al.}{2017a}]{Adams2017a}
Adams, S. M., Kochanek, C. S., Gerke, J. R., \& Stanek, K. Z. 2017a, MNRAS, 469, 1445

\bibitem[\protect\citeauthoryear{Adams et al.}{2017b}]{Adams2017b}
Adams, S. M., Kochanek, C. S., Gerke, J. R., Stanek, K. Z., \& Dai, X. 2017b, MNRAS, 468, 4968

\bibitem[\protect\citeauthoryear{Annis et al.}{2016}]{Annis2016}
Annis, J., Soares-Santos, M., Berger, E., et al. 2016, ApJ, 823, L34

\bibitem[\protect\citeauthoryear{Barron et al.}{2008}]{Barron2008}
Barron, J.T., Stumm C., Hogg, D.W., Lang D. \& Roweis S., 2008, ApJ, 135, 414

\bibitem[\protect\citeauthoryear{Bastian et al.}{2010}]{Bastian2010}
Bastian N.,  Covey K. R.,  Meyer M. R.. , ARA\&A , 2010, vol. 48 pg. 339 

\bibitem[\protect\citeauthoryear{Bond et al.}{2009}]{Bond2009}
Bond, H. E., Bedin, L. R., Bonanos, A. Z., et al. 2009, ApJL, 695, L154

\bibitem[\protect\citeauthoryear{Carrigan}{2009}]{Carrigan2009}
Carrigan R,A, 2009, ApJ, 698, 2075

\bibitem[\protect\citeauthoryear{Chambers}{2016}]{Chambers}
Chambers K.C., Magnier E.A., Metcalfe N. et al., 2016, arXiv:1612.05560

\bibitem[\protect\citeauthoryear{Clayton}{2012}]{Clayton2012}
Clayton, G. C. 2012, Journal of the American Association of Variable Star 
Observers (JAAVSO), 40, 539

\bibitem[\protect\citeauthoryear{Davenport}{2019}]{Davenport2019}
Davenport J., 2019, arXiv: 1907.04443.pdf

\bibitem[\protect\citeauthoryear{Diehl et al.}{2006}]{Diehl2006}
Diehl R., et al., 2006, Nature, 439, 45

\bibitem[\protect\citeauthoryear{Djorgovski et al.}{1998}]{DPOSS}
Djorgovski S.G., Gal R.R., Odewahn  S. C., de Carvalho, R.R., Brunner R., Longo G. \& Scaramella R., 1998, Wide Field Surveys in Cosmology, eds. S. Colombi and Y. Mellier

\bibitem[\protect\citeauthoryear{Djorgovski}{2000}]{Djorgovski2000}
Djorgovski S.G., 2000, A new Era in Bioastronomy, ASP Conference Series, vol 213

\bibitem[\protect\citeauthoryear{Djorgovski et al.}{2001}]{Djorgovski2001}
Djorgovski S.G. et al., 2001, Virtual Observatories of the Future, eds. R. Brunner, S.G. Djorgovski, and A. Szalay, ASP Conference Series, vol. 225

\bibitem[\protect\citeauthoryear{Djorgovski et al.}{2012}]{Djorgovski2012}
Djorgovski S.G. et al., 2012, in Mihara T., Serino M., eds, The First Year of MAXI: Monitoring Variable X-ray Sources, Special Publ. IPCR-127, 263, RIKEN, Tokyo

\bibitem[\protect\citeauthoryear{Drake et al.}{2009}]{Drake2009}
Drake A.J., Djorgovski S.G., Mahabal A. et al., 2009, Astrophysical Journal, 696, 870

\bibitem[\protect\citeauthoryear{Drake et al.}{2019}]{Drake2019}
Drake A.J., Djorgovski S.G., Graham M.J., Stern D., Mahabal A.A., Catelan M., Christensen E., Larson S., 2019, Monthly Notices of the Royal Astronomical Society, 482, 98

\bibitem[\protect\citeauthoryear{Dyson}{1960}]{Dyson1960}
Dyson F., 1960, Science, 131, 1667

\bibitem[\protect\citeauthoryear{Ertl et al.}{2016}]{Ertl2016}
Ertl T., Janka H.-T., Woosley S. E., Sukhbold T., Ugliano M., 2016, ApJ, 818, 124

\bibitem[\protect\citeauthoryear{Flewelling et al.}{2016}]{Flewelling2016}
Flewelling, H. A., Magnier, E. A., Chambers, K. C., et al. 2016, ArXiv e-prints, arXiv:1612.05243

\bibitem[\protect\citeauthoryear{Gaia Collaboration et al.}{2016}]{Gaia2016}
Gaia Collaboration et al., 2018, Astronomy \& Astrophysics, 595, A1

\bibitem[\protect\citeauthoryear{Gaia Collaboration et al.}{2018}]{Gaia2018}
Gaia Collaboration et al., 2018, Astronomy \& Astrophysics, 616, A11


\bibitem[\protect\citeauthoryear{Graham et al.}{2017}]{Graham2017}
Graham  M.J., Djorgovski, S.G., Drake A.J., Stern D., Mahabal A.A., Glikman  E., Larson S. \& Christensen E., 2017,  Monthly Notices of the Royal Astronomical Society, 470, 4112.

\bibitem[\protect\citeauthoryear{Graham et al.}{2019}]{Graham2019}
Graham M.J., Ross N.P., Stern D. et al., 2019, arXiv: 1905.02262 

\bibitem[\protect\citeauthoryear{Greiner et al.}{1990}]{Greiner}
Greiner J., Wenzel W. \& Degel J., 1990, A\&A, 234, 251

\bibitem[\protect\citeauthoryear{Gerke et al.}{2014}]{Gerke2014}
Gerke J.R., Kochanek C.S., Stanek K.Z., 2014, MNRAS, 450, 3289

\bibitem[\protect\citeauthoryear{Griffith et al.}{2015}]{Griffith2015}
Griffith, R. L., Wright, J. T., Maldonado, J., Povich, M.S., Sigurdsson, S. \& Mullan B., 2015, ApJ Supplement Series, 217, 25

\bibitem[\protect\citeauthoryear{Grindlay et al.}{2012}]{Grindlay2012}
Grindlay J., Tang S., Los E., Servillat M., 2012, New Horizons in Time-Domain Astronomy, Proceedings of the International Astronomical Union, IAU Symposium, 285, 29

\bibitem[\protect\citeauthoryear{Hippke et al.}{2015}]{Hippke2015}
Hippke M., Learned J.G., Zee A., Edmondson W.H., Lindner J.F., Kia B., Ditto W.L., \& Stevens I.R., 2015, ApJ, 798, 42

\bibitem[\protect\citeauthoryear{Hoenig et al.}{2014}]{Hoenig2014}
Hoenig S.F., Watson D., Kishimoto M. \& Hjorth J., 2014, Nature, 515, 528

\bibitem[\protect\citeauthoryear{Howard et al.}{2004}]{Howard2004}
Howard, A.W., Horowitz P., Wilkinson D.T., 2004, ApJ, 613, 1270

\bibitem[\protect\citeauthoryear{Howard et al.}{2007}]{Howard2007}
Howard, Andrew, Mead C., Sreetharan P. et al., 2007, Acta Astronautica, 61, 1-6,78

\bibitem[\protect\citeauthoryear{Kaiser et al.}{2002}]{Kaiser2002}
Kaiser, N., Aussel, H., Burke, B. E., et al. 2002, in Society of
Photo-Optical Instrumentation Engineers (SPIE) Conference Series, Vol. 4836, Survey and Other Telescope Technologies and Discoveries, ed. J. A. Tyson \& S. Wolff, 154

\bibitem[\protect\citeauthoryear{Kankare et al.}{2017}]{Kankare2017}
Kankare E., Kotak R., Mattila S. et al., 2017, Nature Astronomy, 1, 865

\bibitem[\protect\citeauthoryear{Kardashev}{1964}]{Kardashev}
Kardashev, N.S., 1964, Soviet Astronomy, 8, 217

\bibitem[\protect\citeauthoryear{Kochanek et al.}{2008}]{Kochanek}
Kochanek, C.S, Beacom, J.F., Kistler, M.D., Prieto, J.L., Stanek, K.Z., Thompson, T.A \& Yuksel H., 2008, ApJ, 684, 1336

\bibitem[\protect\citeauthoryear{Lacki}{2016}]{Lacki2016}
Lacki, B.C., arXiv:1604.07844

\bibitem[\protect\citeauthoryear{Lawrence et al.}{2016}]{Lawrence}
Lawrence et al. 2016, MNRAS, 463, 296

\bibitem[\protect\citeauthoryear{Limongi \& Chieffi.}{2006}]{Limongi2006}
Limongi M., \& Chieffi A. 2006, ApJ, 647, 483

\bibitem[\protect\citeauthoryear{Lingam \& Loeb}{2017}]{Loeb2017}
Lingam M. \& Loeb A., 2017, ApJL, 837, L23

\bibitem[\protect\citeauthoryear{Mack et al.}{2009}]{Mack2009}
Mack K.H., Prieto M.A., Brunetti G. \& Orienti M., 2009, 392, 705

\bibitem[\protect\citeauthoryear{Madsen \& Gaensler}{2013}]{Gaensler}
Madsen G.J. \& B.M. Gaensler, 2013, ApJ Supplement Series, 209, 33

\bibitem[\protect\citeauthoryear{Magnier et al.}{2016a}]{MagnierA}
Magnier, E. A., Schlafly, E. F., Finkbeiner, D. P., et al. 2016a, arXiv:1612.05242

\bibitem[\protect\citeauthoryear{Magnier et al.}{2016b}]{MagnierB}
Magnier, E. A., Sweeney, W. E., Chambers, K. C., et al. 2016b, arXiv:1612.05244

\bibitem[\protect\citeauthoryear{Magnier et al.}{2016c}]{MagnierC}
Magnier, E. A., Chambers, K. C., Flewelling, H. A., et al. 2016c, arXiv:1612.05240

\bibitem[\protect\citeauthoryear{Mahabal et al.}{2009}]{Mahabal2009B}
Mahabal A.A., Djorgovski S.G., Drake A.J. et al., 2009, the Astronomer's Telegram, 2009, Atel: 2029, 2149

\bibitem[\protect\citeauthoryear{Mahabal et al.}{2011}]{Mahabal}
Mahabal A.A., Djorgovski S.G., Drake A.J. et al., 2011, Bull. Astron. Soc. India, 39, 387

\bibitem[\protect\citeauthoryear{Maire et al.}{2018}]{Maire2018}
Maire J., Wright S.A., Cosens M., Antonio F., et al., 2018, Proceedings Volume 10702, Ground-based and Airborne Instrumentation for Astronomy VII; 107025L

\bibitem[\protect\citeauthoryear{Mattsson \& Villarroel}{2017}]{Mattsson2017}
Mattsson L. \& Villarroel B., 2017, Popul\"ar Astronomi, 3, 18

\bibitem[\protect\citeauthoryear{Monet et al.}{2003}]{Monet2003}
Monet, D.G. et al., 2003, ApJ, 125, 984

\bibitem[\protect\citeauthoryear{NASA Technosignature Report}{2018}]{Technosignatures}
N. Participants, NASA Technosignature Report, arXiv: 1812.08681

\bibitem[\protect\citeauthoryear{O'Connor \& Ott}{2011}]{Oconnor2011}
O'Connor, E., \& Ott, C. D. 2011, ApJ, 730, 70

\bibitem[\protect\citeauthoryear{Ott et al.}{2011}]{Ott2011}
Ott, C. D., Reisswig, C., Schnetter, E., et al. 2011, PhRvL, 106, 161103

\bibitem[\protect\citeauthoryear{Prieto}{1997}]{Prieto1997}
Prieto M.A., 1997, MNRAS, 284, 627

%\bibitem[\protect\citeauthoryear{Yasini \& Pelckmans}{2017}]{Pelckmans2017}
%Yasini S. \& Pelckmans K., 2017, IEEE Transactions on Automatic Control, vol. PP, 1

\bibitem[\protect\citeauthoryear{Pelckmans et al.}{in prep.}]{Pelckmans}
Pelckmans K., Laaksoharju M., Villarroel B. et al., in prep.

\bibitem[\protect\citeauthoryear{Robitaille \& Whitney}{2010}]{Robitaille2010}
Robitaille T.~P. \& Whitney B.~A., 2010, ApJL, 710, L11

\bibitem[\protect\citeauthoryear{Rodriguez et al.}{2018}]{Rodriguez2018}
Rodriguez R., Schmidt S.J., Jayasinghe T. et al., 2018, RNAAS, 2, 2

\bibitem[\protect\citeauthoryear{Salpeter}{1955}]{Salpeter1955}
Salpeter E.~E. 1955, ApJ, 121, 161

\bibitem[\protect\citeauthoryear{Scalo}{1986}]{Scalo1986}
Scalo, J. M. 1986, Fundamentals Cosmic Phys., 11, 1

\bibitem[\protect\citeauthoryear{Schwartz \& Townes}{1961}]{Townes}
Schwartz, R. N., \& Townes C.H., 1961, Nature 190, 4772

\bibitem[\protect\citeauthoryear{Shalev-Schwarz \& Ben-David}{2014}]{BenDavid}
Shalev-Schwarz \& Ben-David 2014, ``Understanding Machine Learning: From Theory to Algorithms'', 2014, Cambridge University Press.

\bibitem[\protect\citeauthoryear{Smart et al.}{2009}]{Smart2009}
Smartt S., Eldridge J. J., Crockett R. M., Maund J. R., 2009, MNRAS, 395, 1409

\bibitem[\protect\citeauthoryear{Soodla}{2019}]{Soodla2019}
Soodla J., 2019, ``A System for Cross-matching All-sky Surveys'' (master thesis), 2019, Uppsala Universitet, ISSN 1401-5757

\bibitem[\protect\citeauthoryear{Samus et al.}{2017}]{Sternberg}
Samus N.N., Kazarovets E.V., Durlevich O.V., Kireeva N.N., Pastukhova E.N., 2017, Astronomy Reports, 61, 80

\bibitem[\protect\citeauthoryear{Tang et al.}{2010}]{Tang2010}
Tang, S., Grindlay, J., Los, E., \& Laycock, S. 2010, ApJ, 710, L77

\bibitem[\protect\citeauthoryear{Tang et al.}{2011}]{Tang2011}
Tang, S., Grindlay, J., Los, E., \& Servillat, M. 2011, ApJ, 738, 7

\bibitem[\protect\citeauthoryear{Tang et al.}{2012}]{Tang2012}
Tang, S., Grindlay, J., Moe M., Orosz J.A., Kurucz R.L., Quinn S.N., \& Servillat M., 2012, ApJ, 751, 99

\bibitem[\protect\citeauthoryear{Tellis \& Marcy}{2015}]{Marcy2015}
Tellis, N.K. \& Marcy G.W, 2015., PASP, 127, 952

\bibitem[\protect\citeauthoryear{Tellis \& Marcy}{2017}]{Marcy2017}

\bibitem[\protect\citeauthoryear{Tisserand et al.}{2018}]{Tisserand2018}
Tisserand P., Clayton G.C., Bessell M.S., Welch D.L., Kamath D., Wood P.R., Wils P., Wyrzykowski L., Mroz P. \& Udalski A., 2018, https://arxiv.org/abs/1809.01743

Tellis, N.K. \& Marcy G.W, 2017, ApJ, 153, 251

\bibitem[\protect\citeauthoryear{Ugliano et al.}{2012}]{Ugliano2012}
Ugliano M., Janka H.-T., Marek A., Arcones A., 2012, ApJ , 757, 69

\bibitem[\protect\citeauthoryear{Vavilova et al.}{2012}]{Vavilova2012}
Vavilova, I.B., Pakulyak, L.K., Shlyapnikov, A.A. et al., 2012, Kinematics and Physics of Celestial Bodies, 28, 85

\bibitem[\protect\citeauthoryear{Vavilova}{2016}]{Vavilova2016}
Vavilova, I.B., 2016, Odessa Astronomical Publications, 29, 109

\bibitem[\protect\citeauthoryear{Villarroel et al.}{2016}]{Villarroel2016}
Villarroel B., Imaz I. \& Bergstedt J., 2016, AJ, 152, 76

\bibitem[\protect\citeauthoryear{Waters et al.}{2016}]{Waters2016}
Waters, C. Z., Magnier, E. A., Price, P. A., et al. 2016, ArXiv
e-prints, arXiv:1612.05245

\bibitem[\protect\citeauthoryear{Woosley \& Heger}{2007}]{Woosley2007}
Woosley S. E., \& Heger A. 2007, Phys. Rep., 442, 269

\bibitem[\protect\citeauthoryear{Wright et al.}{2014}]{Wright2014}
Wright, J.T., Mullan B., Sigurdsson, S. \& Povich M.S., 2014, ApJ, 792, 27

\bibitem[\protect\citeauthoryear{Wright et al.}{2018}]{Wright2018}
Wright S.A., Horowitz P., Maire J., Werthimer D., Franklin A., et al., 2018, Proceedings Volume 10702, Ground-based and Airborne Instrumentation for Astronomy VII; 107025I.

\bibitem[\protect\citeauthoryear{Wright et al.}{2014}]{WrightNIR}
Wright S.A., Werthimer D., Treffers R.R. et al., 2014, Proceedings of the SPIE, 9147, 91470J

\bibitem[\protect\citeauthoryear{York et al.}{2000}]{York2000}
York, D.G. et al., 2000, AJ, 120, 1579

\bibitem[\protect\citeauthoryear{Zackrisson et al.}{2015}]{Zackrisson2015}
Zackrisson E., Calissendorff P., Asadi S. \& Nyholm A., 2015, ApJ, 810, 23

\end{thebibliography}


%%%%%%%%%%%%%%%%%%%%%%%%%%%%%%%%%%%%%%%%%%%%%%%%%%

%%%%%%%%%%%%%%%%% APPENDICES %%%%%%%%%%%%%%%%%%%%%

%%\section{Some extra material}

%%TBD

\appendix

\section{Estimated failed supernova rates}\label{sec:supernova}
%In this Section, Lars makes a splendid theoretical demonstration that unless one assumes a crazy and extreme IMF, the number of failed supernovae in the entire Milky Way (400 billion of stars) is maximum one during 50 years... or something like that.   :-)\\

Failed supernovae (SNe) happen when massive stars do not explode as SNe, but rather collapse into black holes without any preceding transient event \citep[see,e.g.][]{Kochanek,Smart2009,Ugliano2012,Ertl2016,Adams2016,Adams2017a,Adams2017b}. The concept is mainly theoretical, as no failed SN has been found observationally with absolute certainty, although there exists at least one good candidate \citep{Adams2016}. Despite the scarcity of empirical evidence, there is still reason to believe that there is a deficit of higher mass SN progenitors as identified by \citet{Kochanek}. But how likely is it that such a failed SN even occurs in the Galaxy within the VASCO timeframe? 

To assess the probability of detecting a failed SN we need to consider two things: (1) the formation rate of massive stars in the Galaxy and (2) which stars (i.e. stars of what mass range) are the progenitors of the failed SNe. The first can be calculated from the total star formation rate for a given stellar initial mass function (IMF). The latter can be estimated from theoretical SN models, but unfortunately these models do not agree very well regarding complete fallback and direct black-hole formation \citep[see, e.g.,][for further details]{Oconnor2011}.

Most observational estimates of the total Galactic star-formation rate (SFR) made over the last four decades fall in the range 0.5-8 $M_{\sun}$~yr$^{-1}$, depending on assumptions about the IMF \citep[see, e.g.,][and references therein]{Diehl2006, Robitaille2010}. Here, we will use the value of \citet{Diehl2006}, namely $\dot{M}_{\rm SFR} = 3.8\,M_{\sun}$~yr$^{-1}$, which is almost exactly in the middle of the above range.  This is based on the \citet{Scalo1986} IMF which has a power-law tail in the high-mass end with slope $-2.7$. In the following, we shall treat as more of less free parameters the high-mass slope, the fraction $f_{\rm LIMS}$ of low- and intermatiate-mass stars (LIMS),and the fraction $f_{\rm HMS}$ of high-mass stars (HMS). We will use the fractions $f_{\rm LIMS}$ and $f_{\rm HMS}= 1- f_{\rm LIMS}$ interchangeably.

On the $\sim$ 100 yr timescale of the VASCO project it is fair to assume a constant galactic-average SFR. Thus, the expected SN rate $\mathcal{R}_{\rm SN}$ is given by:
\begin{equation}
\mathcal{R}_{\rm SN} = \dot{M}_{\rm SFR} \times \int_{m_{\rm SN}}^{m_{\rm max}} \phi(m)\,dm,
\end{equation}
with the IMF expressed as
\begin{equation}
\phi(m,m_0,\epsilon) = \phi_0\,\left({m\over m_0} \right)^{-(1+\epsilon)}\,f_{\rm LIMS}(m_0,m_{\rm SN},\epsilon),
\end{equation}
where $\phi_0$ is a normalization constant chosen such that the average mass $\langle m\rangle = 1$, $m_{\rm SN}$ is the lower mass limit for SNe, $m_0$ is the turn-over mass, i.e., the stellar mass where the IMF peaks and then turns over, and $\epsilon$ is the logarithmic slope of the high-mass end. The LIMS fraction $f_{\rm LIMS}$ is then also a function such that
\begin{equation}
f_{\rm LIMS} = \int_{m_{\rm min}}^{m_{\rm SN}} \phi(m,m_0,\epsilon)\,dm \Bigg/ \,\int_{m_{\rm min}}^{m_{\rm max}} \phi(m,m_0,\epsilon)\,dm.
\end{equation}
The canonical \citet{Salpeter1955}  IMF corresponds to $\epsilon = 1.35$ and $f_{\rm HMS} = 0.003$. To calculate the rate of failed SNe, we simply need to replace the SN mass domain $\mathcal{M}_{\rm SN} \equiv [m_{\rm SN},m_{\rm max}]$ with the corresponding mass domain for failed SNe, $\mathcal{M}_{\rm FSN} $. We consider two estimates of $\mathcal{M}_{\rm FSN} $, based on the results of \citet[][henceforth LC06]{Limongi2006} and \citet[][henceforth WH07]{Woosley2007}, which were chosen because the fraction of failed SNe in these studies is high enough (also at solar metallicity) to have a notable effect on the SN rate \citep[see also][]{Oconnor2011}. 

\begin{figure*}
 \centering
 \resizebox{\hsize}{!}{
   \includegraphics{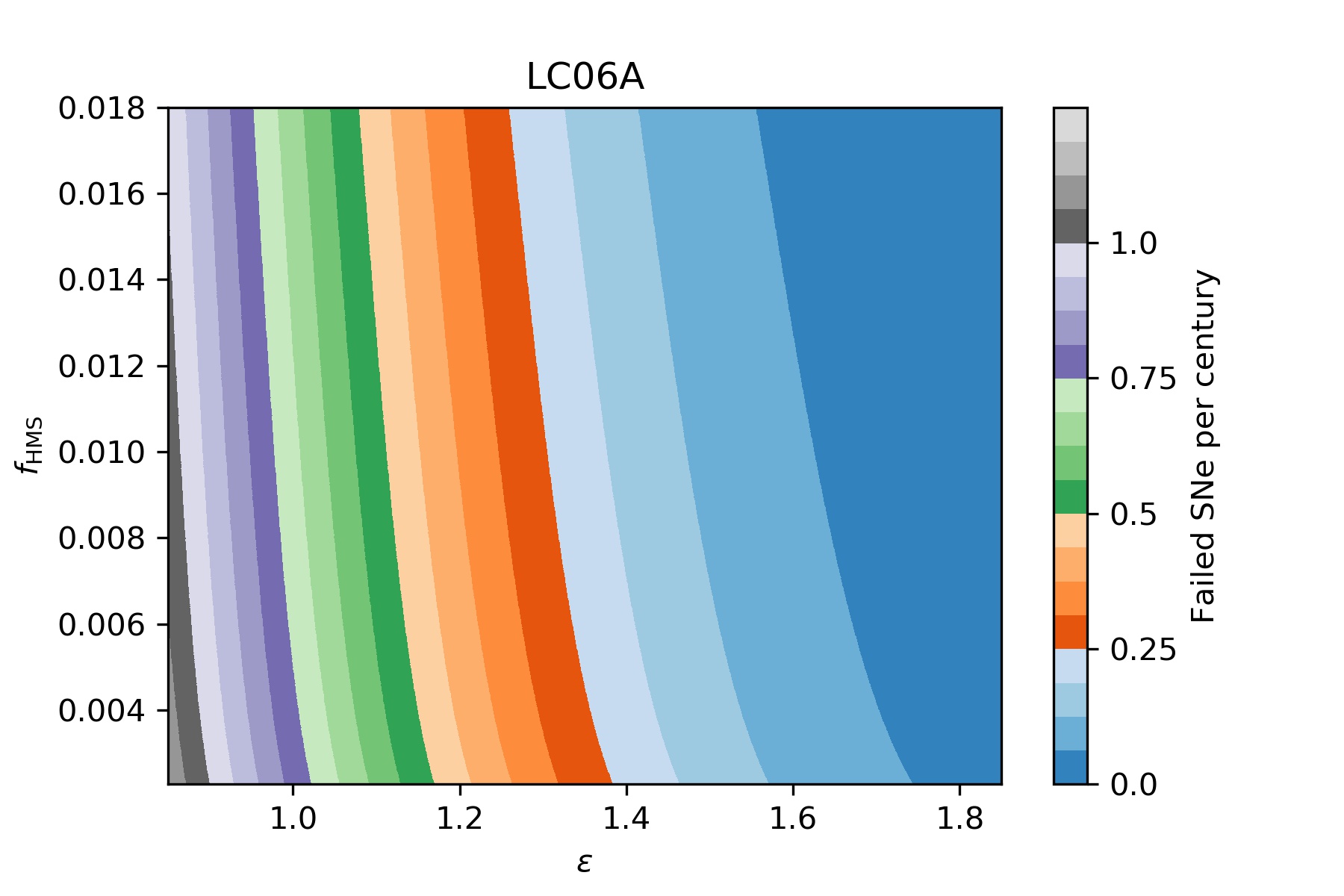}
   \includegraphics{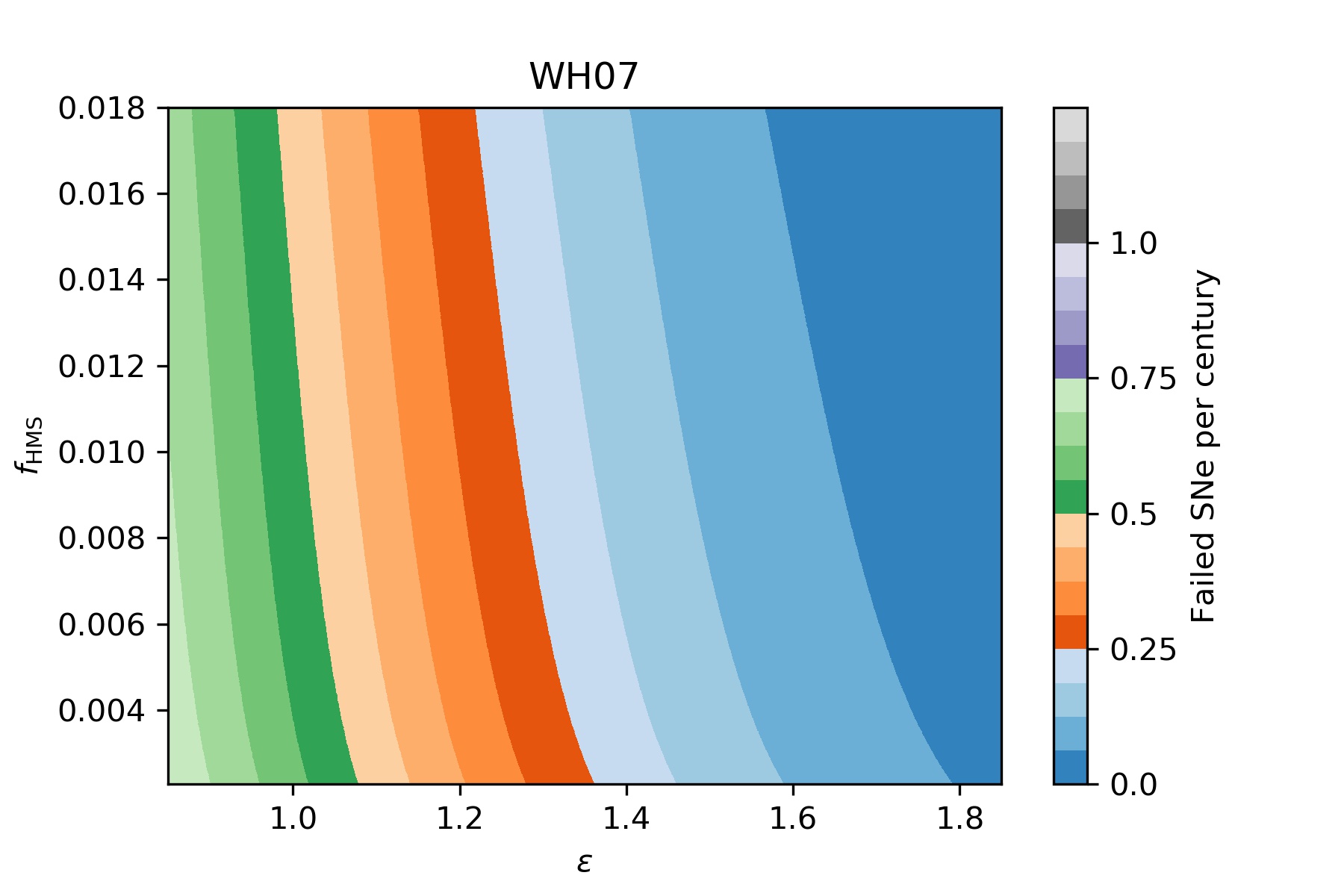}
   }
   \caption{Number of failed SNe per century as a function of the logarithmic slope of the IMF $\epsilon$ and the high-mass fraction $f_{\rm HMS}$ based on SN models by LC06 (left panel) and WH07 (right panel).}
   \label{FailedSN} 
\end{figure*}

Based upon the simple SN-rate model above, we have explored how the rate of failed SNe depends on $\epsilon$ and $f_{\rm LIMS}$ or $f_{\rm HMS}$ for the two estimates of $\mathcal{M}_{\rm FSN} $. The results are shown in Fig. \ref{FailedSN}, from which it is clear that only a top-heavy IMF can produce a rate high enough for it to be likely that we find failed SNe within VASCO. For both the LC06 and the WH07 case an IMF slope $\epsilon = 1.35$ indicate that no failed SN should be detected over a 50 yr time scale ($\mathcal{R}_{\rm FSN} < 0.005$~yr$^{-1}$), for all reasonable values of $f_{\rm LIMS}$ (or $f_{\rm HMS}$). In order to reach two SNe per century, or above, we have to invoke an extremely top-heavy present-day IMF. Essentially, all recent observational constraints on the Galactic IMF indicate an IMF that is actually steeper than the \citet{Salpeter1955} IMF ($\epsilon > 1.35$) and with no strong evidence for variation \citep[see][and references therein]{Bastian2010}, suggesting it is unlikely that we would catch a failed SN in the Galaxy, although it cannot be ruled out.

Our choice of the LC06 and WH07 results as a basis for calculating the rate of failed SNe is somewhat arbitrary, but also conservative. It must be emphasized that it is not well known how common failed SNe actually are. A mass range $[18\,M_{\sun}, 25\,M_{\sun}]$ is sometimes quoted \citep[e.g.,][]{Smart2009,Adams2017a} and theoretical work does indeed seem to lend some support for it \citep{Oconnor2011,Ugliano2012,Ertl2016}. For a normal IMF ($\epsilon=1.35$), $\mathcal{M}_{\rm FSN} \equiv [18\,M_{\sun}, 25\,M_{\sun}]$ leads to a rate $\mathcal{R}_{\rm FSN}\sim 0.025$~yr$^{-1}$. In such a case, the prospect of finding a failed SN within a 50 yr window is somewhat better. Assuming that no stars in the initial-mass range $[18\,M_{\sun}, 25\,M_{\sun}]$ will explode is perhaps too extreme. 

In summary, ``failed SNe", massive stars that collapse to black holes without any detectable transient event (SN explosion) are not likely to explain vanishing stars in the Galaxy on timescales less than $\sim 1000$~yr. Our current understanding of the progenitors of failed SNe (such as the models of LC06 and WH07) indicates that such events should be caught by VASCO only if the present day IMF is extremely top heavy. On a more speculative note, failed SNe are not well understood. Therefore, we cannot rule out the possibility that less massive (and therefore more abundant) stars fail to explode, possibly due to some other mechanism, thus leading to a much higher rate of ``vanishing stars''.

%%%%%%%%%%%%%%%%%%%%%%%%%%%%%%%%%%%%%%%%%%%%%%%%%%

\label{lastpage}
\end{document}